\documentstyle[12pt]{article}
\newcommand{\nc}{\newcommand}
\nc{\noi}{\noindent}  \nc{\Nrm}{{\rm N}}

\nc{\erm}{{\rm e}}  \nc{\Rcal}{{\cal R}}
\nc{\AR}{A_{\rm 1R}} \nc{\ARAR}{A_{\rm 2R}}
\nc{\AT}{A_{\rm 1T}} \nc{\ATAT}{A_{\rm 2T}}
\nc{\al}{\alpha_1} \nc{\alal}{\alpha_2}
\nc{\be}{\beta_1} \nc{\bebe}{\beta_2}
\nc{\psiuno}{\psi_{\rm I}} \nc{\psidue}{\psi_{\rm II}}
\nc{\psitre}{\psi_{\rm III}} \nc{\psiquattro}{\psi_{\rm IV}}
\nc{\psicinque}{\psi_{\rm V}}    \nc{\srm}{{\rm s}}
\nc{\bb}{\begin{equation}} \nc{\ee}{\end{equation}}
\nc{\um}{{1\over 2}} \nc{\C}{I\!\!\!C} \nc{\R}{I\!\!R}
\nc{\pa}{\partial} \nc{\ug}{\; = \;}
\nc{\cent}{\centerline}
\newcommand{\h}{\hspace*{3 ex}}

\newcommand{\Ga}{\Gamma}

\newcommand{\ugg}{ \ = \ }
\newcommand{\ove}{\overline}

\newcommand{\arm}{{\rm a}}
\newcommand{\brm}{{\rm b}}
\newcommand{\crm}{{\rm c}}
\newcommand{\minrm}{{\rm min}}
\newcommand{\rrm}{{\rm r}}
\newcommand{\drm}{{\rm d}}
\newcommand{\eff}{{\rm eff}}
\newcommand{\intrm}{{\rm int}}
\newcommand{\inrm}{{\rm in}}
\newcommand{\A}{{\rm A}}

\newcommand{\Phrm}{{\rm Ph}}
\newcommand{\Rerm}{{\rm Re}}

\newcommand{\Rrm}{{\rm R}}
\newcommand{\Trm}{{\rm T}}
\newcommand{\Drm}{{\rm D}}

\begin{document}

\cent{{\Large {\bf Unified Time Analysis of Photon and}}}

\

\cent{{\Large {\bf (Nonrelativistic) Particle Tunnelling,}}}

\

\cent{{\Large {\bf and the Superluminal group-velocity problem$^{\dagger}$}}}

\footnotetext{$^{\dagger}$ Work partially supported by MURST, INFN, CNR and
by I.N.P./PAN/Krakow ;\\
{\em e-mail addresses\/}: recami@mi.infn.it ; \ olkhovsk@stenos.kiev.ua ; \
jakiel@alf.ifj.edu.pl}

\

\

\cent{Vladislav S.Olkhovsky$^{\arm}$, Erasmo Recami$^{\brm}$ and Jacek
Jakiel$^{\crm}$}

\

\cent{\em {$^{\arm}$ Institute for Nuclear Research, Kiev-03028;
Research Centre "Vidhuk", Kiev, Ukraine.}}
\cent{{\em $^{\brm}$ Facolt\`{a} di Ingegneria, Universit\`{a} statale di Bergamo,
24044 Dalmine (BG), Italy;}}
\cent{{\em I.N.F.N, Sezione di Milano, Milan, Italy; {\rm and}}}
\cent{{\em D.M.O./FEEC and C.C.S., UNICAMP, Campinas, SP, Brazil.}}
\cent{{\em $^{\crm}$ Institute of Nuclear Physics, 31-342 Krak\'{o}w, Poland.}}

\

\

\begin{abstract} A unified approach to the time analysis of tunnelling of
nonrelativistic particles is presented, in which Time is regarded as a
quantum-mechanical observable, canonically conjugated to Energy. The validity
of the Hartman effect (independence of the Tunnelling Time of the opaque
barrier width, with Superluminal group velocities as a consequence) is
verified for {\em all} the known expressions of the mean tunnelling time. \
Moreover, the analogy between particle and photon tunnelling is suitably
exploited. On the basis of such an  analogy, an explanation of some recent
microwave and optics experimental results on tunnelling times is proposed. \
Attention is devoted to some aspects of the causality problem for particle
and photon tunnelling.
\end{abstract}

\vfill\newpage

\section{Introduction.}   %%% Sect.1

\h The study of tunnelling started with the discovery of $\alpha$-decay,
which was followed by Lord Rutherford's investigations and, in 1928,  by
Gamow's quantum-mechanical description[1].  Much later, from the Fifties
onwards, tunnelling experiments in solid state physics (such as those
with tunnelling junctions[2], tunnelling diodes[3] and tunnelling
microscopes[4]) were performed and theoretically analyzed.

\h The study of the {\em tunnelling times} has a long history too.
The problem of the definition of a tunnelling time was mentioned at the
beginning of the Thirties[5,6]. Then, it remained almost ignored until the
Fifties or Sixties, when it was faced the more general question of defining
a quantum-collision duration[7-18]: a question that, in its turn, had been
put aside since the Twenties, after Pauli's works[19] stressing the
impossibility of introducing a self-adjoint operator for Time in quantum
mechanics.  Among the first attempts to regard time as a quantum-mechanical
observable, let us recall refs.[20-26] and, in particular, the
clarification that such a problem received during the Seventies and
Eighties in refs.[27-29].  Reviews about time as a quantum observable
(which results to be a maximal hermitian operator, even if it is not
selfadjoint) canonically conjugated to energy, can be found in
refs.[29-31].  \ [Let us mention that a series of new papers recently
appeared[32-39, and refs. therein], examining the properties of the time
operator in quantum mechanics: however, all such papers seem to ignore the
Naimark theorem[40] which is on the contrary an essential mathematical
basis for refs.[27-31].]
\footnote{{The Naimark theorem (1940) states that a non-orthogonal spectral
decomposition of a maximal hermitian operator can be approximated, with a
weak {\em convergence}, by an orthogonal spectral decomposition with any
desired accuracy degree.}}

\h Recently, developments in various fields of physics and especially the
advent of high-speed electronic devices, based on tunnelling processes,
revived the interest in the tunnelling time analysis, whose relevance had
been previously apparent in nuclear physics only ($\alpha$-radioactivity and,
afterwards, nuclear sub-barrier fission, fusion, proton-radioactivity, etc.).
So that, in recent years, a number of theoretical reviews appeared[41-49].
With regard to experiments on tunnelling times, the great
difficulty with actual measurements for particles was due
to the too small values of the related tunnelling times (see, for instance,
refs.[50-55]). Till when the {\em simulation} of particle tunnelling by
microwave and laser-light tunnelling allowed some very interesting
measurements[56-60] of ``tunnelling times"; such a simulation being based on
the known mathematical analogy between particle and photon tunnelling: Which
becomes evident when
comparing[61-65] the stationary Schroedinger equation, in presence of a
barrier, with the stationary Helmholtz equation for an electromagnetic
wavepacket in a waveguide.

\h In the more interesting {\em time-dependent} case, however, the Schroedinger
and Helmholtz equations are no longer identical: a problem that was left
open, and that one cannot forget. \ Another question that has to be faced
is the introduction of an operator for Time in quantum mechanics and in
quantum electrodynamics. \ Below (in Sects.9-10) we shall tackle with such
problems, as well as with the physical interpretation of some photon
tunnelling experiments.

\h Returning to the question of the theoretical definition of the tunnelling
time for particles, there is not yet a general agreement about such a
definition[41-49]; some reasons being the following: \ (i) The
problem of defining tunnelling times is closely connected with the
more general definition of the quantum-collision duration, and therefore
with the fundamental fact that Time in some cases is just a parameter (like
$x$), but in some other cases is a {\em (quantum) physical observable} (like
$\hat x$); \ (ii) The motion of particles inside a potential barrier is
a quantum phenomenon, that till now has been devoid of any direct classical
limit; \  (iii) There are essential {\em differences} among the initial,
boundary and external conditions assumed within the various definitions
proposed in the literature; those differences have not been analyzed yet.

\h Following ref.[49], we can divide the existing approaches into a few groups,
based ---respectively--- on: \ 1) a time-dependent description in terms of
wavepackets; \ 2) averages over a set of kinematical paths, whose
distribution is supposed to describe the particle motion inside a barrier; \
3) the introduction of a new degree of freedom, constituting a physical clock
for the measurements of tunnelling times.  \ Separately, it stands by itself
the dwell time approach. The latter has {\em ab initio} the presumptive
meaning of the time during which the incident flux has to be maintained,
to provide the accumulated particle storage in the barrier[9,49]. \
The {\em first group} contains the so-called phase times (firstly mentioned in
[7,8] and applied to tunnelling in refs.[66,67]), the times related to the motion
of the wavepacket spatial centroid (considered for generic quantum
collisions in refs.[17,18] and in particular for tunnelling in [68,69]), and
finally the Olkhovsky--Recami (OR) method[48,70,71] (based on the
generalization of the time durations defined for atomic and nuclear
collisions in refs.[11,29,30]), which adopts averages over fluxes pointing in
a well-defined direction {\em only}, and has recourse to a quantum operator
for Time.

\h The {\em second group} contains methods utilizing the Feynman
path integrals[72-75], the Wigner distribution paths[76,77], and the Bohm
approach[78].

\h The methods with a Larmor clock[79] or an oscillating barrier[80,81]
pertain to the {\em third group}.

\h In our opinion, basic self-consistent definitions of tunnelling durations
(mean values, variances, and so on) should be worked out in
a way similar to the one followed when defining in quantum mechanics other
physical quantities (like distances, energies, momenta, etc.): namely, by
utilizing the properties of time as a quantum observable.  One ought then
choose a time {\em operator}, canonically conjugated to the energy operator;
and take advantage of the equivalence between the averages performed in the
time and in the energy representation,\footnote{{An equivalence still
following from the Naimark[40] theorem!}}
with adequate weights (measures). \
For such definitions, it is obviously necessary to abandon any descriptions
in terms of plane waves, and to have rather recourse to wavepackets and to
the time-dependent Schroedinger equation: As it is actually typical in the
quantum collision theory (see, e.g., the third one of refs.[10]). \
Afterwards, one will finally operate within the framework of conventional
quantum mechanics; and, within this framework, it will be possible to show
(as we shall do) that every known definition of tunnelling time is
---at least in some suitable asymptotic regions--- either a {\em particular
case} of the most general definition, or a definition valid (not for
tunnelling but) for some accompanying process.

\h The necessary formalism, and the consequent definitions, will be introduced
in Sect.2 below. \ In Sects.3$\div$5 we shall briefly compare one another the
various existing approaches. \ In Sect.6 we shall discuss some peculiarities
of the tunnelling evolution. \ In Sect.7 we shall show the Hartman--Fletcher
effect to be valid for all the known expressions of the mean tunnelling
times. \ The short Sect.8 will present a new ``two-phase description" of
tunnelling, which is convenient for media without absorption and dissipation,
as well as for Josephson junctions. \ In Sect.9 we investigate the analogies
between the (time-dependent) Schroedinger equation, in presence of a quantum
barrier, and the (time-dependent) Helmholtz equation for an electromagnetic
wavepacket in a waveguide, and discuss the ``tunnelling" experiments with
microwaves. \ In Sect.10 we go on to study the tunnelling times in the optical
``tunnelling" experiments based on frustrated total internal reflection. \
In Sect.11, a short note follows on the reshaping (and reconstruction)
phenomena, in connection with a possible formulation of the principle of
``relativistic causality" which is valid also when the tunnelling velocities
are actually Superluminal. \ Finally, in Sect.12, some conclusions are
presented, together with some prospective considerations for the near future.

\section{A quantum operator for Time as the starting point for defining
the tunnelling durations. \ The OR formalism.}      %%%Sect.2

\h We confine ourselves to the simple case of particles moving only along
the $x$-direction, and consider a time-independent barrier located in the
interval ($0,a$): \ See Fig.1, in which a larger interval, ($x_{i}, x_{f}$),
containing the barrier region, is also indicated. \ [We shall call
{\em region} II the barrier region, {\em region} I the (initial) one on its
left, and {\em region} III the (final) one on its right]. \ Following the
known definition of duration of a collision ---set forth firstly in ref.[11],
then in [27,29,30] and afterwards generalized in [48,71]--- we can
eventually define the mean value $\langle t_{\pm}(x) \rangle$ of the time $t$
at which a particle passes through position $x$ ({\em travelling in the
positive or negative direction, respectively\/}:

\begin{equation}  \langle t_{\pm}(x) \rangle  = \frac{\int_{-\infty}^{\infty}
t J_{\pm}(x,t)\drm t}{\int_{-\infty}^{\infty} J_{\pm}(x,t)\drm t}  \end{equation}

\noi and the variance $\Drm \equiv \sigma^2$ of that time distribution:

\begin{equation} \Drm \, t_{\pm}(x) = \frac{\int_{-\infty}^{\infty} t^{2}
J_{\pm}(x,t)\drm t}{\int_{-\infty}^{\infty} J_{\pm}(x,t)\drm t} -[\langle t_{\pm}(x)
\rangle]^{2} \ , \end{equation}

\noi $J_{\pm}(x,t)$ representing the {\em positive} or {\em negative values},
respectively, of the probability flux density $J(x,t)={\rm Re}[(i \hbar/m)
\Psi (x,t) \partial \Psi^{\dagger}(x,t)/\partial x]$ of a wavepacket
$\Psi (x,t)$ evolving in time;\footnote{{Let us mention that one could
measure the quantities $J_{\pm}$,
at least in principle, via the following experimental set-up:  (i) for measuring
$J_+$, one can have recourse to two detectors, the first one measuring the
incident flux $J_{\inrm}$, while the second one  ---sufficiently far away,
but still located before the barrier--- measures the same incoming flux (at
the new position) in delayed coincidence with the former measurement; \
analogously, (ii) to measure $J_-$, the first detector will measure the
reflected flux $J_\Rrm$, while the second one measures the same (reflected)
flux in {\em advanced} coincidence with the former.}}
\ namely \ $J_{\pm}(x,t) \equiv J \, \Theta({\pm} J)$. \ Let us repeat that,
with appropriate averaging weights, the (canonically conjugate) time and
energy representations are equivalent in the sense that: \ $\langle ...
\rangle_{t} = \langle ...\rangle_{E}$. \ Below, for the sake of simplicity,
we shall omit the index $t$ in all expressions for $\langle ...
\rangle_{t}$. \ Let us also re-emphasize that the mentioned equivalence is a
consequence of the existence in quantum mechanics of a unique operator for
time: which, even if not self-adjoint (i.e., with a uniquely defined but
non-orthogonal spectral decomposition)[19,20], is however (maximal)
{\em hermitian}; it is represented by the time variable $t$ in the
$t-$representation for square-integrable space-time wavepackets, and, in
the case of a continuum energy spectrum, by $-i \hbar \partial / \partial
E$ in the $E-$representation, for the Fourier-transformed wavepackets
(provided that point $E=0$ is eliminated[29b], i.e.
for wavepackets with moving back-tails and, of course, nonzero
fluxes; one can notice that states with zero energy E would not play any role,
anyway, in collision experiments).[27-31]\\

\h Let us stress that this Olkhovsky-Recami (OR) approach is just a direct
consequence of conventional
quantum mechanics. From the ordinary probabilistic interpretation of $\rho
(x,t)$ and from the well-known continuity equation

$$ {\partial{\rho(x,t)} \over {\partial t}} + {\pa J(x,t) \over \pa x} = 0
\ ,$$

it  follows {\em also in this (more general) case} that the two weights
$w_+$ and $w_-$

$$w_{+}(x,t) = J_{+}(x,t) \; \left[ \int_{-\infty}^{\infty} J_{+}(x,t)
 \: \drm t  \right]^{-1}$$

\

$$w_{-}(x,t) = J_{-}(x,t) \; \left[ \int_{-\infty}^{\infty} J_{-}(x,t)
 \: \drm t \right]^{-1}  \; ,$$

can be regarded as the probabilities that our ``particle" passes through
position $x$ during a unit time--interval centred at $t$ (in the case of
forward and backward motion, respectively).

\h Actually, for those time intervals for which $J =
J_{+}$ or $J = J_{-}$, one can rewrite the continuity equation as follows:

$${{\partial {\rho_{>}(x,t)} \over \partial t} =
- {\pa J_{+}(x,t) \over {\pa x}}} $$

\

$${{\partial {\rho_{<}(x,t)} \over \partial t} = -{\pa J_{-}(x,t) \over
\pa x} \ ,}$$

\

respectively. \ These relations can be considered  as formal
definitions of \ $\partial {\rho_{>}} / \partial t$ and \ $\partial
{\rho_{<}} / \partial {t}$. \ \ Let us now integrate them
over time from $-\infty$ to $t$; \ we obtain:

$$\rho_{>}(x,t)= -\int_{-\infty}^{t}
{\pa J_{+}(x,{t}') \over \pa x} \: \drm {t}' $$

$$\rho_{<}(x,t)= -\int_{-\infty}^{t}
 {\pa J_{-}(x,{t}') \over \pa x} \: \drm t' $$

\

with the initial conditions \ $\rho_{>}(x,-\infty)=\rho_{<}(x,-\infty)=0$.
 \ Then, let us introduce the quantities

$$N_{>} (x,\infty;t) \equiv \int_{x}^{\infty} \rho_{>}({x}',t) \, \drm {x}'
= \int_{-\infty}^{t} J_{+}(x,{t'}) \, \drm {t}' \ >0 $$

\

$$N_{<} (-\infty,x;t) \equiv \int_{-\infty}^{x} \rho_{<}({x}',t) \,
\drm {x}' =
-\int_{-\infty}^{t} J_{-}(x,{t}') \, \drm {t}' \ >0 \ , $$

\

which have  the meaning of probabilities for our ``particle" to be located
at time $t$ on the semi-axis  $(x,\infty)$ or $(-\infty,x)$
respectively, as functions of the flux densities $J_{+}(x,t)$ or
$J_{-}(x,t)$, provided that the normalization condition \
$\int_{-\infty}^{\infty}\rho(x,t) \drm x = 1$ \ is fulfilled.  \
The r.h.s.'s of the last couple of equations have been obtained by integrating
the r.h.s.'s of the above expressions for $\rho_{>}(x,t)$ and  $\rho_{<}(x,t)$
and by adopting the boundary conditions  \
$J_{+}(-\infty,t) = J_{-}(-\infty,t) = 0$. \  Now, by
differentiating $N_{>} (x,\infty;t)$ and $N_{<} (-\infty,x;t)$ with respect
to $t$, one obtains:

\
$${{\partial{N_{>}}(x,\infty,t) \over \partial{t}} =
J_{+}(x,t)  > 0 } $$

\

$${{\partial{N_{<}}(x,-\infty,t) \over \partial{t}} =
- \, J_{-}(x,t)  > 0 } \ . $$

{\em Finally}, from our last four equations one can infer that:

$${w_{+}(x,t) ={{\partial {N_{>}}(x,\infty;t)/\partial {t}
\over {N_{>}(x,-\infty;\infty)}}}} $$

$${w_{-}(x,t) ={{\partial {N_{<}}(x,-\infty;t)/\partial {t}
\over {N_{<}(-\infty,x;\infty)}}}} \ , $$

which justify the aforementioned probabilistic interpretation of
$w_{+}(x,t)$ and $w_{-}(x,t)$. \ Let us notice, incidentally, that our
approach does {\em not} assume any ad hoc postulate.\\

\h Our previous OR formalism is therefore enough for defining
mean values, variances (and other ``dispersions") related with the duration
distributions of various collisions, including tunnelling. \ For instance,
for transmissions from region I to region III we have

\begin{equation}  \langle \tau_{\rm T}(x_{i},x_{f}) \rangle =\langle
t_{+}(x_{f}) \rangle - \langle t_{+}(x_{i}) \rangle \end{equation}

\begin{equation}  \Drm \, \tau_{\rm T}(x_{i},x_{f}) = \Drm \, t_{+}(x_{f}) +
\Drm \, t_{+}(x_{i}) \end{equation}

\noi with $-\infty < x_{i} \leq 0 $ and $a \leq x_{f} < \infty $. \ For a
mere {\em tunnelling} process, one has

\begin{equation}  \langle t_{\rm tun}(0,a) \rangle =\langle t_{+}(a) \rangle
- \langle t_{+}(0) \rangle \end{equation}

\noi and

\begin{equation}  \Drm \, \tau_{\rm tun}(0,a) = \Drm \, t_{+}(a) + \Drm \,
t_{+}(0) \ . \end{equation}

\noi For penetration (till a point $x_{f}$ inside the barrier region II),
similar expressions hold for $\langle \tau_{\rm pen}(x_{i},x_{f}) \rangle$
and $\Drm \, \tau_{\rm pen}(x_{i},x_{f})$,  with $ 0 < x_{f} <  a$.  \ For
reflections at generic points, located in regions I or II, with
$x_i \leq x_{f}  < a$,  one has

\begin{equation}  \langle \tau_{\rm R}(x_{i},x_{f}) \rangle =\langle
t_{-}(x_{f}) \rangle - \langle t_{+}(x_{i}) \rangle \end{equation}

\noi and

\begin{equation}  \Drm \, \tau_{R}(x_{i},x_{f}) = \Drm \, t_{-}(x_{f}) + \Drm \,
t_{+}(x_{i}) \end{equation}

\h Let us repeat that these definitions hold within the framework of
conventional quantum mechanics, {\em without} introducing any new physical
postulates.

\h In the asymptotic cases, when $|x_{i}| >> a$, it is:

$$\langle \tau_{\rm T}^{\rm as}(x_{i},x_{f}) \rangle =\langle t(x_{f})
\rangle_{\rm T} - \langle t(x_{i}) \rangle_{\rm in} \eqno(3a)$$

\noi and

\begin{equation}  \langle \tau_{\rm T}^{\rm as}(x_{i},x_{f}) \rangle =
\langle \tau_{\rm T}(x_{i},x_{f}) \rangle + \langle t_{+}(x_{i}) \rangle -
\langle t(x_{i}) \rangle_{\rm in} \end{equation}

\noi where  $\langle...\rangle_{\rm T}$ and $\langle...\rangle_{\rm in}$
are averages over the fluxes corresponding to $\psi_{\rm T} = A_{\rm T}
\exp (ikx)$ \ and \ to $\psi_{\rm in} = \exp (ikx)$, respectively. \ For
initial wavepackets of the form

\[ \Psi (x,t)= \int_{0}^{\infty} G(k-\ove{k})\exp{(i(kx-Et)/\hbar)}\drm k \]

\noi (where $E=\hbar^{2}k^{2}/2m$; \ $\int_{0}^{\infty} | G(k-\ove{k})
|^{2}\drm E = 1$; \ $G(0)=G(\infty) = 0$; \ $ k>0$; \ $\ove{k}$ being the
value corresponding to the peak\footnote{{For real
tunnelling, with under-barrier energies, one should actually multiply the
weight amplitude $G$ by a cutoff function, which in the case of a rectangular
barrier with height $V_0$ is simply $\Theta (E - V_0)$.}}  of $G$)
and for sufficiently small energy (or momentum) spreads, when

\[  \int_{0}^{\infty} \upsilon^{n} | GA_{\rm T}|^{2}\drm E \approx
\int_{0}^{\infty} \upsilon^{n} | G |^{2} \drm E \]

\noi with $n=0,1; \ \upsilon \equiv \hbar k/m$, one gets:

\begin{equation}  \langle \tau_{\rm T}^{\rm as}(x_{i},x_{f}) \rangle =
{\langle \tau_{\rm T}^{\Phrm}(x_{i},x_{f}) \rangle}_E \ , \end{equation}

\noi where

\[ \langle ...\rangle_{E}= \int_{0}^{\infty}\drm E \upsilon
| G(k-\ove{k}) |^{2}... / \int_{0}^{\infty} \upsilon | G |^{2} \drm E \ . \]

\noi The quantity

\begin{equation}   \tau_{\rm T}^{\rm Ph}(x_{i},x_{f}) = (1/\upsilon)(x_{f}
- x_{i}) + \hbar \, \drm (arg A_{\rm T})/\drm E    \end{equation}

\noi is the transmission phase time obtained by the stationary-phase
approximation.
 \ In the same approximation, and when it is small the contribution of
$\Drm \, t_{+}(x_{i})$ to the variance $\Drm \, \tau_{\rm T} (x_{i},x_{f})$
(that can be realized for sufficiently {\em large} energy spreads, i.e. for
spatially short wavepackets), we obtain:

\begin{equation}
\Drm \, \tau_{\rm T}(x_{i},x_{f})= \hbar^{2} [\langle (\partial | A_{\rm T} |
 / \partial E)^{2} \rangle_{E} / \langle | A_{\rm T} | ^2 \rangle_{E}
\end{equation}

\noi In the opposite case of very {\em small} energy spreads, i.e.,
quasi-monochromatic particles, the expression (12) becomes just that
part of $\Drm \, t_{+}(x_{f})$ and  $\Drm \, t_{\rm T}(x_{i} ,x_{f})$  which
is due to the barrier presence.

\h  In the limit $| G |^{2} \rightarrow \delta(E-\ove{E})$, when it is
$\ove{E} \equiv \hbar^{2} \ove{k}^{2}/2m$, the equation (10) does yield the
ordinary phase time, without averaging. \ For a rectangular barrier with
height $V_{0}$ and $\kappa a >>1$ \ (where $\kappa \equiv [2m(V_{0} -E
)]^{1/2}/
\hbar)$, \ the expressions (10) and (12) for $x_{i}=0$ and $x_{f}=a$
transform, in the same limit, into the well-known expressions

$$\tau_{\rm tun}^{\rm Ph} \rightarrow   \frac{2}{\upsilon k} \eqno(11a)$$

\noi (see ref.[66], and also [48,49]), and

$$(\Drm \, \tau_{\rm tun}^{\rm Ph})^{1/2} \rightarrow  \frac{ak}{\upsilon}
 \ , \eqno(12a)$$       

\noi respectively. \ It should be noticed that our eq.(12a) {\em coincides}
with one of the Larmor times[79] and with the B\"{u}ttiker-Landauer time[80], as
well as with the imaginary part of the complex time in the Feynman
path-integral approach (see also ref.[82]).

\h Recently G.Nimtz stressed the
importance of the simple relation (11a), that he heuristically verified, and
called it a ``universal property" of tunnelling times. \ Actually, eqs.(11a)
and (12a) can strongly help clarifying many of the current discussions
about tunnelling times. \ Let us add, incidentally, that recent theoretical
work by Abolhasani and Golshani[71], which regards the OR
approach as giving the most natural definition for a transmission time
within the standard interpretation of quantum mechanics, conludes that
the best times that could be obtained in Bohmian mechanics are {\em the
same} as OR's.

\h For a real weight amplitude $G(k-\ove{k})$, when $ \langle t(0)
\rangle >_{\rm in}=0 $, \ from (9) we obtain

\begin{equation}     \langle \tau_{\rm tun} (0,a) \rangle = \langle
\tau_{\rm tun}^{\rm Ph} \rangle - \langle t_{+} (0) \rangle \ . \end{equation}

\h By the way, if the experimental conditions are such that only the
positive-momentum components of the wavepackets are recorded, i.e., \
$  \Lambda_{\exp,+} \Psi (x_{i},.t) = \Psi_{\rm in} (x_{i},.t)$, \
quantity $\Lambda_{\exp,+}$ being the projector onto the positive-momentum
states, then for any $x_{i}$ in the range $(-\infty,0)$ and $x_{f}$ in the
range $(a,\infty)$ it will be:

$$\langle \tau_{\rm T}(x_{i},x_{f}) \rangle_{\exp} = \langle
\tau_{\rm T}^{\rm Ph}(x_{i},x_{f}) \rangle_{E} \eqno(10a)$$

\noi and

$$\langle \tau_{\rm tun}(0,a) \rangle_{\exp} = \langle
\tau_{\rm tun}^{\rm Ph} \rangle_{E} \ , \eqno(13a)$$

\noi since $\langle t(0) \rangle_{\exp} = \langle t(0) \rangle_{\rm in}$.

\h The main criticism, by the authors of refs.[49,64] and also [78,83], of any
approach to the definition of tunnelling times in which a spatial or temporal
averaging over moving wavepackets is adopted, invokes the lack of a causal
relationship between the incoming peak or ``centroid" and the outgoing peak
or ``centroid". \ It was already clear in the Sixties (see, for instance,
ref.[18]) that such criticism is valid only when finite (not asymptotic)
distances from the interaction region are considered. \ Moreover, that
criticism applies more to attempts like the one in ref.[69] (where it was
looked for the evolution of an incoming into an outgoing peak), than for our
definitions of collision, tunnelling, transmission, penetration, reflection
(etc.) durations: In fact, our definitions for the mean duration of any such
processes do {\em not} assume that the centroid (or peak) of the incident
wavepacket directly evolves into the centroid (or peak) of the transmitted
and reflected packets. \ Our definitions are simply {\em differences}
between the mean times referring to the passage of the final and initial
wavepackets through the relevant space-points, regardless of any intermediate
motion, transformation or reshaping of those wavepackets... \ At last, for
each collision (etc.) process as a whole, we shall be able to test the
causality condition.

\h Actually, there is no a single general formulation of the causality
condition, which be necessary and sufficient for all possible cases of
collisions (both for nonrelativistic and relativistic wavepackets). \
The simplest (or strongest) nonrelativistic condition implies the
non-negativity of the mean durations. This is, however, a sufficient but not
necessary causality condition.\footnote{{In fact, let us recall that: \ (i)
all the ordinary causal paradoxes seem to be solvable[84] within Special
Relativity, when it is {\em not} restricted to subluminal motions only; \
(ii) nevertheless, whenever it is met an object ${\cal O}$
travelling at Superluminal speed, negative contributions ought to be expected
to the tunnelling times[85]: and this should not to be regarded as
unphysical[84]. \ In fact, whenever the {\em object} ${\cal O}$
{\em overcomes} the infinite speed with respect to a certain observer,
it will afterwards appear to the same observer as its {\em anti}-object
$\ove{\cal O}$ travelling in the opposite {\em space} direction[84]. \
For instance, when going on from the lab to a frame ${\cal F}$ moving in the
{\em same} direction as the particles or waves entering the barrier region, the
objects $\cal O$ penetrating through the final part of the barrier (with
almost infinite speed[86]) will appear in the frame ${\cal F}$ as
anti-objects $\ove{\cal O}$ crossing that portion of the barrier {\em in
the opposite space--direction}[84].  In the new frame ${\cal F}$,
therefore, such anti-objects $\ove {\cal O}$ would yield a {\em negative}
contribution to the tunnelling time: which could even result, in total, to
be negative. \ What we want to stress here is that the appearance
of such negative times is predicted by Relativity itself, on the basis of
the ordinary postulates[84-86]. \ From the theoretical point of view,
besides refs.[85,86,84], see also refs.[87]. \  From the (quite interesting!)
experimental point of view, see refs.[88].}}  \ Negative times
(advance phenomena) were revealed even near nuclear resonances, distorted
by the nonresonant background (see, in particular, ref.[30]); similarly,
``advance" phenomena can occur also at the beginning of tunnelling (see Sect.6
below).

\h Generally speaking, a complete causality condition should be connected not
only with the mean time duration, but also with other temporal properties
of the considered process. For example, the following variant could seem to be
more realistic: $<<$The difference $t^\eff_\A(x_i,x_f) = t^\eff_f - t^\eff_i$ ,
between the effective arrival-instant of the flux at $x_f$ and the effective
start-instant of the flux at $x_i$, is to be non-negative (where A = T, pen,
tun,...)$>>$; where the effective instants are defined as  \ $t^\eff_f \equiv
\langle t(x_{f}) \rangle + \sigma[t(x_{f})]$, \ and \ $t^\eff_i \equiv
\langle t(x_{i}) \rangle - \sigma[t(x_{i})]$, \ the standard deviations being
of course \ $\sigma[t(x_{f})] = [\Drm \, t(x_f)]^{1/2}$; \
$\sigma[t(x_{i})] = [\Drm \, t(x_i)]^{1/2}$; \ so that:

\

$$t^\eff_\A(x_i,x_f) \equiv t^\eff_f - t^\eff_i \ugg \langle t(x_{f}) \rangle
- \langle t(x_{i}) \rangle + \sigma[t(x_{f})] + \sigma[t(x_{i})] \ .$$

\

\noi But this condition too is sufficient but not necessary, because often
wavepackets are represented with infinite and not very rapidly decreasing
forward-tails... More realistic formulations of the causality condition for
wavepackets (with very long tails) will be presented in Sect.8.

\section{The meaning of the mean dwell time.}
%%%   [Sect.3]

\h  As it is known[89] (see also ref.[71]), the mean dwell time can be
presented in two equivalent forms:

\begin{equation} \langle \tau^{\rm Dw}(x_{i},x_{f}) \rangle =
\frac{\int_{-\infty}^{\infty}\drm t \int_{x_{i}}^{x_{f}} | \Psi(x,t)|^{2}
\drm x}{ \int_{-\infty}^{\infty} J_{\rm in}(x_{i},t) \drm t} \end{equation}

\noi and

$$\langle \tau^{\rm Dw}(x_{i},x_{f}) \rangle = \frac{\int_{-\infty}^{\infty}
t J(x_{f},t)\drm t-\int_{-\infty}^{\infty}t J(x_{i},t)\drm t}{ \int_{-\infty}^{\infty}
J_{\rm in}(x_{i},t) \drm t} \ , \eqno(14')$$

\noi with $-\infty < x_{i}  \leq 0 \;$ and $\; a \leq x_{f}  < \infty$. \ Let
us observe that in the first definition, eq.(14), of the mean dwell time,
in integrating over $t$ it is used a weight different from the one introduced
by us in Sect.2. \ Let us comment on the meaning of the weight function
(the ``measure"). Taking into account the relation
$\int_{-\infty}^{\infty} J_{\rm in}(x_{i},t)\drm t=
\int_{-\infty}^{\infty} | \Psi(x,t)|^{2} \drm x $, which follows from the
continuity equation, one can easily see that the weight of eq.(14)
is \ $dP(x,t) =
{| \Psi(x,t)|^{2} \drm x} \; / \; {\int_{-\infty}^{\infty} | \Psi(x,t)
|^{2} \drm x} \;$, \ which has the well-known quantum-mechanical meaning of
probability for a particle to be {\em localized}, or to
{\em dwell}, in the spatial region $(x,x+\drm x)$ at the instant $t$,
independently of the motion direction. \ Then, the integrated quantity \
$P(x_{1},x_{2};t) = {\int_{x_{1}}^{x_{2}} | \Psi(x,t)|^{2} \drm x}
\; / \; {\int_{-\infty}^{\infty} | \Psi(x,t) |^{2} \drm x} \;$, \
has the meaning of probability of finding the particle inside the spatial
interval $(x_{i} ,x_{f})$ at the instant $t$ (see also ref.[90]).

\h The equivalence of relations (14) and (14') is a consequence of the
continuity equation which links the probabilities associated with the two
processes: ``dwelling inside" and ``passing through" the interval
$(x_{i} ,x_{f})$. \ However, we can note that the applicability of the
integrated weight $P(x_{1} ,x_{2} ;t)$ for the time analysis (in contrast
with the space analysis) is limited, since it allows calculating the
mean dwell times only, but not their variances.

\h Taking into account that \ $J(x_{i},t)=J_{\rm in}(x_{i},t) +
J_{\rm R}(x_{i},t) + J_{\rm int}(x_{i},t)$ \ and \ $J(x_{f},t) =
J_{\rm T}(x_{f},t)$ \ (where $J_{\rm in}$, $J_{\rm R}$ and $J_{\rm T}$
correspond to the wavepackets $\Psi_{\rm in}(x_{i},t)$, $\Psi_{R}(x_{i},t)$
and $\Psi_{\rm T}(x_{f},t)$, which have been constructed in terms of the
stationary wave functions $\psi_{\rm in}$, $\psi_{\rm R} =A_{\rm R} \exp
(-ikx)$ and $\psi_{\rm T}$, respectively), \ and that for $J_\intrm$ (originating
from the interference between $\Psi_\inrm(x_i,t)$ and $\Psi_\Rrm(x_i,t)$)
it holds

 \[  J_{\rm int}(x,t)=\Rerm{(i\hbar /m)[\Psi_{\rm in}(x,t) \; \partial
 \Psi_{\rm R}^{*}(x,t)/\partial x + \Psi_{\rm R}(x,t) \; \partial
 \Psi_{\rm in}^{*}(x,t)/ \partial x]}  \]

\noi and

          \[  \int_{-\infty}^{\infty} J_{\rm int}(x_{i},t) \drm t=0  \]

\noi we eventually obtain the interesting relation

\begin{equation}  \langle \tau^{\rm Dw}(x_{i} ,x_{f}) \rangle = \langle T
\rangle_{E}  \langle \tau_{\rm T} (x_{i} ,x_{f}) \rangle +  \langle R(x_{i})
\rangle_{E}  \langle \tau_{\rm R} (x_{i} ,x_{f}) \rangle  \end{equation}

\noi with  $\langle T \rangle_{E} = \langle |A_{\rm T}|^2  \upsilon
\rangle_{E} / \langle \upsilon \rangle_{E}$, \ $\langle R(x_{i})
\rangle_{E} = \langle R \rangle_{E} + \langle r(x_{i}) \rangle$, \
$\langle R \rangle_{E} = \langle |A_{\rm R}|^2 \upsilon \rangle_{E} /
\langle \upsilon \rangle_{E}$, \ $\langle T \rangle_{E} + \langle R
\rangle_{E} =1$, \ and with

\[  \langle r(x) \rangle = \frac{\int_{-\infty}^{\infty} [J_{+}(x,t) -
J_{\rm in}(x,t)]\drm t}{ \int_{-\infty}^{\infty} J_{\rm in}(x_{i},t) \drm t}
 \ . \]

\noi We stress that $ \langle r(x) \rangle $ is negative and tends to $0$
when $x$ tends to $-\infty$.

\h   When  $\Psi_{\rm in}(x_{i},t)$ and $\Psi_{R} (x_{i},t)$ are well
separated in time, i.e. $\langle r(x) \rangle =0$, one obtains the simple
well-known[33] {\em weighted average rule:}

\begin{equation}  \langle \tau^{\rm Dw}(x_{i} ,x_{f}) \rangle = \langle T
\rangle_{E} \; \langle \tau_{T} (x_{i} ,x_{f}) \rangle + \langle R
\rangle_{E} \; \langle \tau_{\rm R} (x_{i} ,x_{f}) \rangle  \end{equation}

\h For a rectangular barrier with $\kappa a >> 1$  and quasi-monochromatic
particles, the expressions (15) and (16) with $x_{i} =0$ and $x_{f} =a$
transform into the known expressions

$$ \langle \tau^{\rm Dw}(x_{i} ,x_{f}) \rangle = \langle \hbar k/\kappa V_{0}
\rangle_{E} \ , \eqno(15a)$$

\noi (were we took account of the interference term  $\langle r(x) \rangle $),
and

$$ \langle \tau^{\rm Dw}(x_{i} ,x_{f}) \rangle = \langle 2/\kappa \upsilon
\rangle_{E} \ , \eqno(16a)$$

\noi (where the interference term $\langle r(x) \rangle $ is now equal to $0$).

\h  When $A_{\rm R} = 0$, i.e. the barrier is transparent, the mean dwell
time (14),(14') is automatically equal to

\begin{equation}  \langle \tau^{\rm Dw}(x_{i} ,x_{f}) \rangle = \langle
\tau_{\rm T}(x_{i} ,x_{f}) \rangle \ . \end{equation}

\h It is not clear, however, how to define {\em directly} the variances of
the dwell-time distributions. The approach proposed in ref.[91] seems rather
artificial, with its abrupt switching on of the initial wavepacket. \
It is possible to define the variances of the dwell-time distributions
{\em indirectly}, for example by means of relation (15), when basing
ourselves on the standard deviations $\sigma(\tau_\Trm)$, $\sigma(\tau_\Rrm)$
of the transmission-time and reflection-time distributions.

\section{A brief analysis of the Larmor and B\"{u}ttiker-Landauer ``clock" approaches}
%%%Sect.4

\h One can realize that the introduction of additional degrees of freedom as
``clocks" may distort the true values of the tunnelling time. The Larmor
clock uses the phenomenon of the change of the spin orientation
(the Larmor precession and spin-flip) in a weak homogeneous magnetic field
superposed to the barrier region. \ If initially the particle is polarized
in the  $x$ direction, after the tunnelling its spin gets small $y$  and
$z$ components. The Larmor times  $\tau_{y,{\rm T}}^{\rm La}$ and
$\tau_{z,{\rm T}}^{\rm La}$ are defined by the ratio of the spin-rotation
angles [on their turn, defined by the $y-$ and $z-$ spin components] to the
(precession and rotation) frequency[13,14,79]. \ For an opaque rectangular
barrier with $\kappa a >> 1$, the two expressions were obtained:

\begin{equation} \langle \tau_{y,{\rm tun}}^{\rm La} \rangle = \langle
\tau^{\rm Dw}(x_{i} ,x_{f}) \rangle = \langle \hbar k/\kappa V_{0}
\rangle_{E} \end{equation}

\noi and

\begin{equation} \langle \tau_{z,{\rm tun}}^{\rm La} \rangle =\langle ma/
\hbar k \rangle_{E} \ . \end{equation}

\h In refs.[48,82] it was noted that, if the magnetic field region is
infinitely extended, the expression (18) just yields
---after having averaged over the small energy spread of the
wavepacket--- the phase tunnelling time, eq.(11a).

\h As to eq.(19), it refers in reality not to a rotation, but to a {\em jump}
to ``spin-up" or ``spin-down" (spin-flip), together with a Zeeman
energy-level splitting[49,79]. \ Due to the Zeeman splitting, the spin
component parallel to the magnetic field corresponds to a higher
tunnelling energy, and hence the particle tunnels preferentially to that
state.  This explains why the tunnelling time $\tau_{z,{\rm tun}}^{\rm La}$
entering eq.(19) depends only on the absolute value $|A_{\rm T}|$ (or rather
on $\drm |A_\Trm| / \drm E$), and coincides with expression (12a).

\h The B\"{u}ttiker-Landauer clock[49,80,81] is connected with the oscillation
of the barrier (absorption and emission of ``modulation" quanta), during
tunnelling.  Also in this case one can realize (for the same reasons as for
$ \langle \tau_{z,{\rm tun}}^{\rm La} \rangle$) that the coincidence of the
B\"{u}ttiker-Landauer time with eq.(12a) is connected with the energy
dependence of $|A_{\rm T}|$.

\section{A short analysis of the kinematical-path approaches}
%%%  Sect.5

\h The Feynman path-integral approach to quantum mechanics was applied in
[72-75] to evaluate the mean tunnelling time (by averaging over all the
paths that have the same beginning and end points) with the complex weight
factor $\exp [iS(x(t))/\hbar]$, where $S$ is the action associated with
the path $x(t)$. \ Such a weighting of the tunnelling times implies the
appearance of real and imaginary components[49].  In ref.[72] the real and
imaginary parts of the complex tunnelling time were found to be equal to
$\langle \tau_{y,{\rm tun}}^{\rm La} \rangle$ and to -$ \langle
\tau_{z,{\rm tun}}^{\rm La} \rangle$, respectively. \ An interesting
development of this approach, its instanton version, is presented in ref.[75].
The instanton-bounce path is a stationary point in the Euclidean action integral.
Such a path is obtained by analytic continuation to imaginary time of the
Feynman-path integrands (which contain the factor $\exp (iS/\hbar)$).
This path obeys a classical equation of motion inside the potential barrier
with its sign reversed (so that it actually becomes a well). In ref.[75] the
instanton bounces were considered as real physical processes. The bounce
duration was calculated in real time, and was found to be in good agreement
with the one evaluated via the phase-time method. The temporal density of
bounces was estimated in imaginary time, and the obtained result ---in the
phase-time approximation limit--- coincided with the tunnelling-time standard
deviation (as given by eq.(12)). Here one can see a manifestation of the
equivalence (in the phase-time approximation) of the Schroedinger and
Feynman representations of quantum mechanics.

\h Another definition of the tunnelling time is connected with the Wigner
path distribution[76,77]. The basic idea of this approach, reformulated
by Muga, Brouard and Sala, is that the tunnelling-time distribution
for a wavepacket can be obtained by considering a classical ensemble of
particles with a certain distribution function, namely the Wigner function
$f(x,p)$: so that the flux at position $x$ can be separated into positive
and negative components:

\begin{equation}     J(x) = J^{+}(x)  +  J^{-}(x)  \end{equation}

\noi with $J^{+} = \int^{\infty}_{0}(p/m) f(x,p)dp$ \ and \ $ J^{-}  = J
-J^{+}$. \ They formally obtained the same expressions (3) and (5),
for the transmission, tunnelling and penetration durations, as in the
OR formalism, provided that $J^{\pm}$ replaces our $J_{\pm}$. \ The dwell
time decomposition, then, takes the form

\begin{equation}  \langle \tau^{\rm Dw}(x_{i} ,x_{f}) \rangle = \langle T
\rangle_{E} \; \langle \tau_{\rm T} (x_{i} ,x_{f}) \rangle +
\langle R_{\rm M}(x_{i}) \rangle_{E}  \; \langle \tau_{R} (x_{i} ,x_{f})
\rangle  \end{equation}

\noi with $R_{\rm M}(x) = \int^{\infty}_{0} |J^{-}(x,t)| \drm t$. \ Asymptotically,
$\langle R_{\rm M}(x) \rangle$ tends to our quantity $\langle R \rangle_{E}$
and eq.(21) takes the form of the known ``weighted average rule" (16).

\h One more alternative is the stochastic method for wavepackets in ref.[92].
It also leads to real times, but its numerical implementation is not
trivial[93].

\h In ref.[83] the Bohm approach to quantum mechanics was used to choose a
set of classical paths which do not cross each other. The Bohm formulation,
on one side, can be regarded as equivalent to the Schroedinger
equation[94], while on the other side can perhaps provide a basis for a
nonstandard interpretation of quantum mechanics[49]. The expression obtained
in ref.[83] for the mean dwell time is not only positive definite but also
unambiguously distinguishes between transmitted and reflected particles:

\begin{equation} \tau^{\rm Dw}(x_{i},x_{f})  = \int_{0}^{\infty}\drm t
\int_{x_{i}}^{x_{f}} | \Psi(x,t)|^{2} \drm x = T \, \tau_{\rm T} (x_{i} ,x_{f}) +
R \, \tau_{\rm R} (x_{i} ,x_{f}) \end{equation} %%%eq.(22)

\noi with

\begin{equation} \tau_{\rm T}(x_{i},x_{f})  = \int_{0}^{\infty}\drm t
\int_{x_{i}}^{x_{f}} | \Psi(x,t)|^{2} \Theta (x-x_{\rm c}) \drm x / T  \end{equation}
%%%eq.(23)

\begin{equation} \tau_{\rm R}(x_{i},x_{f})  = \int_{0}^{\infty}\drm t
\int_{x_{i}}^{x_{f}} | \Psi(x,t)|^{2} \Theta (x_{\rm c}-x) \drm x / T  \end{equation}
%%%eq.(24)

\noi where $T$ and $R$ are the mean transmission and reflection probability,
respectively. \ The ``bifurcation line" $x_{\rm c} =x_{\rm c}(t)$,
which separates the transmitted from the reflected trajectories, is defined
through the relation

\begin{equation} T= \int_{-\infty}^{\infty}\drm t | \Psi(x,t)|^{2}
\Theta (x-x_{\rm c}) \drm x \ . \end{equation}  %%%(25)

\noi Let us add that two main differences exist between this (Leavens') and
our formalism: \ (i) a difference in the temporal integrations
(which are $\int^{\infty}_{0}\drm t $  and ($\int^{\infty}_{-\infty}\drm t $,
respectively), that sometimes are relevant; and \ (ii) a difference in
the separation of the fluxes, that we operate ``by sign" (cf. eqs.(1),(2))
and here it is operated by the line $x_{\rm c}$:

\begin{equation}   J(x,t) = [J(x,t)]_{\rm T}  +  [J(x,t)]_{\rm R}
\end{equation}    %%%eq.(26)

\noi with \ $[J(x,t)]_{\rm T} = J(x,t) \; \Theta (x - x_{\rm c}(t))$, \
$[J(x,t)]_{\rm R}= J(x,t) \; \Theta (x_{\rm c}(t) - x)$.

\

\section{Characteristics of the tunnelling evolution}    %%%Sect.6

\h The results of the calculations presented in ref.[71], within the OR
formalism, show that: \ (i) {\em the mean tunnelling time does not depend
on the barrier width $a$ for
sufficiently large $a$} (``Hartman effect"); \ (ii) the quantity
$\langle \tau_{\rm tun}(0,a) \rangle$ decreases when the energy increases;
 \ (iii) the value of $\langle \tau_{\rm pen}(0,x) \rangle$  rapidly
increases for increasing $x$ near $x=0$  and afterwards tends to saturation
(even if with a very slight, continuous increasing) for values near $x=a$;
and \ (iv) at variance with ref.[95], no plot for the mean
penetration time of our wavepackets presents any interval
with negative values\footnote{{See, however, footnote$^{\|)}$}}, nor with
negative slop for increasing $x$.

\h In Fig.2 the dependence of the values of $\langle \tau_{\rm tun}(0,a)
\rangle$ on $a$ is presented for gaussian wave packets

$$G(k- \overline k) \equiv C \exp [{- (k-\overline k)^2} /
{(2 \, \Delta k)^{2}}] $$

\noi and rectangular barriers with the same parameters as in ref.[95]: namely,
$V_{0} =10$ eV; \
$\ove{E}=2.5$, 5, and 7.5 eV with $\Delta k = 0.02 \; {\AA}^{-1}$ (curves 1a, 2a,
3a respectively); \ and $\ove{E} = 5$ eV with $\Delta k = 0.04 \; {\AA}^{-1}$
and $0.06 \; {\AA}^{-1}$ (curves 4a, 5a, respectively). \ On the contrary, the
curves for $\langle t_{+}(a) \rangle$, corresponding to different energies
and different $\Delta k$, are all practically superposed to the single
curve 6. \ Moreover, since $\langle \tau_{\rm tun}^{\rm Ph} \rangle$
depends only very weakly on $a$, the quantity $\langle \tau_{\rm tun}(0,a)
\rangle$ depends on $a$ essentially through the term $\langle t_{+}(0)
\rangle$ (see curves 1b$\div$5b).

\h Let us emphasize that all these calculations show that
$\langle t_{+}(0) \rangle$ assumes {\em negative} values (see also [96]).
Such ``acausal" time-advance is a result
of the interference between the incoming waves and the waves reflected by
the barrier forward edge: It happens that the reflected wavepacket
cancel out the back edge of the incoming-wavepacket, and the larger the
barrier width, the larger is the part of the incoming-wavepacket back edge
which is extinguished by the reflected waves, up to the saturation (when
the contribution of the reflected-wavepacket becomes almost constant,
independently of $a$). \ Since all $\langle t_{+}(0) \rangle$ are
negative, eq.(13) yields that the values of $\langle \tau_{\rm tun}(0,a)
\rangle$ are always positive and larger than $\langle \tau_{\rm tun}^{\rm Ph}
\rangle$. \ In connection with this fact, it is worthwhile to note that the
example with a classical ensemble of two particles (one with a large
above-barrier energy and the other with a small sub-barrier energy),
presented in ref.[93], does not seem to be well-grounded, not only because
that tunnelling is a quantum phenomenon without a direct classical limit,
but, first of all, because ref.[93] overlooks the fact that the values of
$\langle t_{+}(0) \rangle$ are negative.

\h Let us mention that the last calculations by Zaichenko[96] (with the
same parameters) have shown that such time-advance is noticeable also
{\em before} the barrier front (even if {\em near} the barrier front wall,
only). He found also negative values of $\langle \tau_{\rm pen}(x_{i},x_{f})
\rangle$, for instance, for $x_{i} = - a/5$ and $x_{f}$ in the interval $0$
to $2a/5$: but this result too is not acausal, because the last equation of
Sect.2 (for example) is fulfilled in this case.\\

\section{On the general validity of the Hartman--Fletcher effect (HE)}
%%%Sect.7

\h We called[48] ``Hartman--Fletcher effect" (or for simplicity ``Hartman
effect", HE) the fact that for opaque potential barriers the mean tunnelling
time does not depend on the barrier width, so that for large barriers the
effective tunnelling--velocity can become arbitrarily large. \ Such effect
was first studied in refs.[66,67] by the stationary-phase method for the
one-dimensional tunnelling of quasi-monochromatic nonrelativistic particles;
where it was found that the {\em phase tunnelling time}

\begin{equation} \tau_{\rm tun}^{\rm Ph}  = \hbar \; \drm (\arg A_{\rm T}+ka)
/\drm E  \end{equation}          %%%       (27)

\noi (which equals the {\em mean tunnelling time} $\langle \tau_{\rm tun}
\rangle$  when it is possible to neglect the interference between incident
and reflected waves outside the barrier[48]) was independent of $a$. \ In
fact, for a rectangular potential barrier, it holds in particular that \
$A_{\rm T} = 4ik \kappa [(k^{2} -\kappa^{2})\Drm_{-} +2ik \kappa
\Drm_{+}]^{-1} \exp [-(\kappa+ik)a]$, \ with \ $\Drm_{\pm}=1 {\pm}
\exp (-2 \kappa)$, \ and that \ $\tau^{\rm Ph}_{\rm tun}= 2/(\upsilon
\kappa)$ \ when $\kappa a >> 1$.

\h Now we shall test the validity of the HE for all the other theoretical
expressions proposed for the mean tunnelling times. \ Let us first consider
the {\em mean dwell time} $\langle \tau^{\rm Dw}_{\rm tun} \rangle$, ref.[89],
the mean Larmor time $\langle \tau^{\rm La}_{y,{\rm tun}} \rangle$,[79,13]
and {\em the real part} of the complex tunnelling time obtained by averaging
over {\em the Feynman paths} ${\rm Re} \tau^{\rm F}_{\rm tun}$, ref.[72],
which all equal $\hbar k/(\kappa V_{0})$ in the case of quasi-monochromatic
particles and opaque rectangular barriers: One can immediately see[61]
that also in these cases there is {\em no} dependence on
the barrier width, and consequently the HE is valid. \ As to the OR
nonrelativistic approach, developed in refs.[48,71,96], the validity of the
HE for the mean tunnelling time can be inferred directly from the expression

\begin{equation}  \langle t_{\rm tun} \rangle =\langle t_{+}(a) \rangle -
\langle t_{+}(0) \rangle = \langle \tau_{\rm tun}^{\rm Ph} \rangle_{E} -
\langle t_{+}(0) \rangle \ ; \end{equation}         %%%eq.(28)

\noi it was moreover confirmed by the numerous calculations performed in the
same set of papers[48,71,96] for various cases of gaussian wavepackets
(see also Sect.6 above).

\h Let us now consider, by contrast, {\em the second Larmor time\/}[69]

\begin{equation} \tau^{\rm L}_{z,{\rm tun}}= {{\hbar [\langle (\partial
| A_{\rm T} | / \partial E)^{2} \rangle} \over {\langle | A_{\rm T} | ^2
\rangle ]^{1/2}}} \ , \end{equation}                 %%%eq.(29)

\noi the {\em B\"{u}ttiker-Landauer time} $\tau_{\rm tun}^{\rm B-L}$,[80]
 \ and {\em the imaginary part} of the complex tunnelling time
${\rm Im} \tau_{\rm tun}^{\rm F}$,[72]  obtained within the {\em Feynman
approach}, which too are equal to eq.(29): \ They all become equal to \
$a \mu/(\hbar \kappa)$, i.e., they all are {\em proportional} to the barrier
width $a$, in the opaque rectangular-barrier limit;[61] \ so that the
HE is {\em not} valid for them! \ However, it was shown in ref.[48] that
these last three
times are {\em not mean times}, but merely {\em standard deviations} (or ``mean
square fluctuations") of the tunnelling-time distributions, because they are
equal to \ $[\Drm_{\rm dyn} \tau_{\rm tun}]^{1/2}$, \ where \
$\Drm_{\rm dyn} \tau_{\rm tun}$ \ is that part of \ $\Drm \, t_{+}(x_{f})$ \
[or analogously of $\Drm \, t_{\rm T}(x_{i},x_{f})$] \ which is due to the
barrier presence and is defined by the simple equation \ $\Drm_{\rm dyn} \;
\tau_{\rm tun} = \Drm \, \tau_{\rm tun} - \Drm \, t_{+}(0)$, \ where \
$\Drm \tau_{\rm tun} =  \langle  \tau_{\rm tun}^{2} \rangle   -  \langle
\tau_{\rm tun} \rangle  ^{2}$ \ and \ $ \langle  \tau_{\rm tun}^{2} \rangle
=  \langle  [t_{+}(a) - \langle t_{+}(0) \rangle  ]^{2} \rangle   + \Drm \,
t_{+}(0)$. \ In conclusion, the former three times are not connected with
the peak (or group) velocity of the tunnelling particles, but with the
{\em spread} of the tunnelling velocity distributions.

\h All these results are obtained for transparent media (without absorption
and dissipation). As it was theoretically demonstrated in ref.[97] within
nonrelativistic quantum mechanics, the HE vanishes for barriers with high
enough absorption. This was confirmed experimentally for electromagnetic
(microwave) tunnelling in ref.[98].

\h The tunnelling through potential barriers with dissipation will be
examined elsewhere.

\h Here let only add a comment.  From some papers[99], it seems that the
integral
penetration time, needed to cross a portion of a barrier, in the case of a
very long barrier starts to increase again ---after the plateau corresponding
to infinite speed--- proportionally to the distance.  This is due to the
contribution of the above-barrier frequencies (or energies) contained in
the considered wavepackets, which become more and more important as the
tunnelling components are progressively damped down. \ In this paper,
however, we refer to the behaviour of the {\em tunnelling} (or, in the
classical case, of the evanescent) waves.\\

\

\section{The Two-phase description of tunnelling}  %%%Sect.8

\h Let us mention also a new description of tunnelling which can be convenient
for transparent media, and also for Josephson junctions.  In such a
representation the transmission and reflection amplitudes have been
rewritten[100,61] (for the same external boundary conditions in Fig.1)
in the form

\begin{equation} A_{\rm T} = i \, {\rm Im} \, (\exp (i \varphi_{1}))
\exp (i \varphi_{2}-ika) , \ \ A_{\rm R} =  {\rm Re} \, (\exp (i \varphi_{1}))
\exp (i \varphi_{2}-ika) \ , \end{equation}          %%%eq.(30)

\noi where the phases $\varphi_{1}$ and $\varphi_{2}$ are typical parameters
for the description of a two-element monodromic matrix $S$, or of a
two-channel collision matrix $S$; with elements $S_{00}= S_{11} = A_{\rm T}$
and $S_{01} = S_{10} = A_{\rm R}$ and with the unitarity condition [$i,j,k =
0,1$]

$$\Sigma_{j=0}^{1} S_{ij} S^{*}_{jk} \ugg \delta_{ik} \ . $$

\h In particular, for rectangular potential barriers it is \ $\varphi_{1} =
\arctan\{2 \sigma /[(1+ \sigma^{2}) \sinh (\kappa a)]\} $, \ and \
$\varphi_{2} = \arctan\{ \sigma \sinh(\kappa a)/[\sinh^{2} (\kappa a/2) -
\sigma^{2} \cosh^{2} (\kappa a/2)]\}$, \ with \ $\sigma = \kappa /k$ \ and \
$\kappa^{2} = \kappa^{2}_{0} - k^{2}$, \ it being $\kappa_{0} =
[2\mu V_{0}]^{1/2} / \hbar$. \ In terms the the phases $\varphi_{1}$ and
$\varphi_{2}$, the expressions for \ $\tau_{\rm tun}^{\rm Ph}$ \ and \
$\tau_{z,{\rm tun}}^{\rm L} = \tau_{\rm tun}^{\rm B\mbox{--}L}$ \ acquire
the following form:

\begin{equation} \tau_{\rm tun}^{\rm Ph}=\hbar \frac{\partial (\arg A_{\rm T}
 + ka)} {\partial E} = \hbar \frac{\partial (\varphi_{2})}{\partial E} \; ; \ \
\tau_{z,{\rm tun}}^{\rm L}=\tau_{\rm tun}^{\rm B\mbox{--}L} = \hbar
\frac{\partial \varphi_{1}}{\partial E} \, \cot (\varphi_{1}) \ .
\end{equation}    %%%eq.(31)

\noi So, one can see that in the opaque barrier limit the phases
$\varphi_{2}$, or $\varphi_{1}$, enter into the play only when the considered
{\em times} are dependent on $a$, or independent of $a$, respectively.

\h For the times $\langle  \tau_{\rm tun}^{\rm Dw} \rangle = \langle
\tau_{z,{\rm tun}}^{\rm L} \rangle$, one obtains in this formalism a
complicated expression, which can be represented[61] only in terms of
{\em both} $\varphi_{1}$ and $\varphi_{2}$.

\h In the presence of absorption, both phases become complex and hence the
formulae (31) become much more lengthy, and in general depend on $a$
with a violation of the HE, in accordance with refs.[97,98].\\

\section{Time-dependent Scr\"{o}dinger and Helmholtz equations: Similarities and
distinctions between their solutions.}   %%%Sect.9

\h The formal analogy is well-known between the (time-independent)
Schroedinger equation in presence of a potential barrier and the
(time-independent) Helmholtz equation for a wave-guided beam; this was the
basis for regarding the evanescent waves in suitable (``undersized")
waveguides as {\em simulating} the case of tunnelling photons. \ We want here
to study analogies and differences between the corresponding {\em
time-dependent} equations. \ Let us mention, incidentally, that a similar
analysis for the relativistic particle case was performed for instance in
refs.[101,102].

\h Here we shall deal with the comparison of the
{\em solutions} of the time-dependent Schroedinger equation (for
nonrelativistic particles) and of the time-dependent Helmholtz equation for
electromagnetic waves. In the time-dependent case such equations are no
longer mathematically identical, since the time derivative appear at the
fist order in the former and at the second order in the latter. \ We shall
take advantage, however, of a similarity between the probabilistic
interpretation of the wave function for a quantum particle and for a
classical electromagnetic wavepacket (cf., e.g., refs.[103]); this will be
enough for introducing identical definitions of the mean time instants and
durations (and variances, etc.) in the two cases (see also refs.[104,105]).

\h Concretely, let us consider the Helmholtz equation for
the case of an electromagnetic wavepacket in the hollow rectangular waveguide,
with an
``undersized" segment, depicted in Fig.3 (with cross section $a {\times} b$ in
its narrow part, it being $a < b$), which was largely employed in experiments
with microwaves[56].  Inside the waveguide, the time-dependent wave equation
for any of the vector quantities $\vec{A}, \vec{E}, \vec{H}$ is of the type

\begin{equation}  \Delta \vec{A} - \frac{1}{c^{2}}\frac{\partial^{2} \vec{A}}{\partial t^{2}} = 0  \end{equation}
%  (32)

\noi where $\vec{A}$ is the vector potential, with the subsidiary gauge
condition ${\rm div} \vec{A}= 0$, while $\vec{E} =-(1/c) \partial \vec{A}
/ \partial t $ is the electric field strength, and $\vec{H} = {\rm rot}
\vec{A}$ is the magnetic field strength. \ As is known (see, for instance,
refs.[106-108]), for boundary the conditions

\begin{eqnarray}   E_{y} = 0  \ \ {\rm for} \ z = 0 \ {\rm and} \
z = a  \nonumber \\  E_{z} = 0 \ \ {\rm for} \  y = 0 \ {\rm and} \
y = b  \end{eqnarray}   %%%eq.(32)

\noi the monochromatic solution of eq.(32) can be represented as a
superposition of the following waves (for definiteness we chose TE-waves):

\

\begin{eqnarray}    E_{x} = 0  \nonumber \\
E_{y}^{\pm} = E_{0} \sin{(k_{z} z)} \cos{(k_{y} y)} \exp{[i(\omega t {\pm}
\gamma x)]}  \nonumber  \\
E_{z}^{\pm} = -E_{0}(k_{y}/k_{z}) \cos{(k_{z} z)} \sin{(k_{y} y)}
\exp{[i(\omega t {\pm} \gamma x)]} \ ,
\end{eqnarray}

\noi with  $k_{z}^{2} + k_{y}^{2} + \gamma^{2} = \omega^{2} /c^{2} = (2\pi
/\lambda )^{2}$, \ $k_{z} =m \pi /a $, \ $k_{y} =n \pi /b$, \  $m$  and
$n$  being integer numbers. \ Thus:

\begin{equation} \gamma = 2 \pi [(\frac{1}{\lambda})^{2} -
(\frac{1}{\lambda_{c}})^{2}]^{1/2}, \; (\frac{1}{\lambda_{c}})^{2} =
(\frac{m}{2a})^{2}+(\frac{n}{2a})^{2} \end{equation}  %eq.(35)

\noi where $\gamma$ is real ($\gamma ={\rm Re} \; \gamma$) if $ \lambda
< \lambda_{c}$, and $\lambda$ is imaginary ($\lambda ={\rm Im} \;
\lambda$) if  $\lambda > \lambda_{c}$. \ Similar expressions for $\lambda$
were obtained for TH-waves[56,107].

\h Generally speaking, any solution of eq.(32) can be written as a wavepacket
constructed from monochromatic solutions (34), analogously to what hold for
any solution of the time-dependent Schroedinger equation. Without
forgetting that in the first-quantization scheme, a probabilistic
single-photon wave function can be represented[103,109] by a wavepacket for
$\vec{A}$: which in the case of plane waves writes for example

\begin{equation} \vec{A}(\vec{r},t)=\int_{k>0} \frac{\drm^{3}\vec{k}}{k}
\vec{\chi}(\vec{k}) \; \exp{(i\vec{k} {\cdot} \vec{r} - ikct)} \, \end{equation}
%%%eq.(36)

\noi where $\vec{r} \equiv (x, y, z)$; \  $\vec{\chi}(\vec{k}) =
\sum_{i=1}^{2} \chi_{i} (\vec{k}) \; \vec{e_{i}} (\vec{k})$; \
$\vec{e_{i}} {\cdot} \vec{e_{j}} =\delta_{i,j}$; \  $\vec{e_{i}} {\cdot}
\vec{k}=0$; \ $i,j = 1,2 $ (or $i,j = y,z$ if $\vec{k} {\cdot} \vec{r} =
k_{x}x$); \ $k=|\vec{k}|$; \ $k=\omega /c$; \ and \ $\chi_{i} (\vec{k})$
is the amplitude for the photon to have momentum $\vec{k}$ and polarization
$\vec{e_{i}}$, \ so that \ $ |\chi_{i} (\vec{k})|^{2} d\vec{k}$ is then
proportional to the probability that the photon have a momentum between
$\vec{k}$ and $\vec{k} +d\vec{k}$ in the polarization state $\vec{e_{i}}$.
 \ Though it is not possible to localize a photon in the direction of its
polarization, nevertheless, {\em for one-dimensional propagation} it is
possible to use the space-time probabilistic interpretation of eq.(36)
along the axis $x$ (the propagation direction)[109,105]. This can be realized
from the following. Usually one does not have recourse directly to the
probability density and probability flux density, but rather to the
the {\em energy density} $s_{0}$ and the {\em energy flux density}
$s_{x}$; they however do {\em not} constitute a 4-dimensional vector,
being components of the energy-momentum tensor. Only in two dimensions
their continuity equation[103] is Lorentz invariant!; we can write down it
(for {\em one space dimension\/}) as:

\begin{equation}   \partial s_{0} / \partial t  +\partial s_{x} /\partial x
= 0  \ ,  \end{equation}     %%%eq.(37)

\noi where

\begin{equation}    s_{0} =[{\vec{E}}^* {\cdot} \vec{E} +
{\vec{H}}^* {\cdot} \vec{H}]/{8\pi}, \ \  s_{x} = c \; {\rm Re} [{\vec{E}}^*
{\cdot} \vec{H} ]_{x} /{2\pi}  \end{equation}   %%%eq.(38)

\noi and the axis $x$ is the motion direction (i.e., the mean momentum
direction) of the wavepacket (36). \ As a normalization condition one can
identify the integrals over space of $s_{0}$ and $s_{x}$ with the mean
photon energy and the mean photon momentum, respectively.  With this
normalization, which bypasses the problem of the impossibility of a direct
probabilistic interpretation in space of eq.(36), we can define {\em by
convention} as

\begin{equation}  \rho_{\rm em} \, \drm x = {{S_{0} \, \drm x} \over
{\int S_{0} \, \drm x}}, \ \
S_{0} = \int s_{0} \, \drm y \drm z     \end{equation}  %%%eq.(39)

\noi the probability density for a photon to be localized in the one-dimensional
space interval ($x,x+\drm x$) along the axis $x$ at time $t$, and as

\begin{equation}  J_{{\rm em},x} \, \drm x = {{S_{x} \drm x} \over {\int S_{x}
\drm x}}, \ \  S_{x} = \int s_{x} \, \drm y \drm z  \end{equation} %%%eq.(40)

\noi the flux probability density for a photon to pass through point $x$
(i.e., through the plane orthogonal at $x$ to the $x$-axis) during the time
interval ($t,t+\drm t$); \ on the analogy of the probabilistic quantities
ordinarily introduced for particles. \ The justification, and convenience,
of such definitions are also supported by the coincidence of the wavepacket
group velocity with the velocity of the energy transportation, which was
established for electromagnetic plane-waves packets in the vacuum; see,
e.g., ref.[110]. For a definition of group velocity in the case of evanescent
waves, see Appendix B in ref.[111].

\ In conclusion, the solution (36) of the time-dependent Helmholtz equation
(for relativistic electromagnetic wavepackets) is quite similar to the
plane-wave packet solution of the time-dependent Schroedinger equation
(for non-relativistic quantum particles), with the following differences:

(i) the space-time probabilistic interpretation of eq.(36) is valid only
in the one-dimensional space case, at variance with the Schroedinger case.
It is interesting that the same conclusion holds for waveguides or
transparent media, when reflections and tunnellings can take place; in
particular, for the waveguides depicted in Fig.3, and for optical
experiments (with frustrated total reflection)[51,52] in the case, e.g.,
of a double prism arrangement;

(ii) the energy-wavenumber relation for non-relativistic particles (corresponding
to selfadjoint, linear Hamiltonians) is quadratic: for instance, in vacuum
it is $E = \hbar^2 k^2 / 2m$; this leads to the fact that wavepackets  do
always {\em spread}. \ By contrast, the energy-wavenumber relation for
photons in the vacuum is linear: $E = \hbar c k$; and therefore there is
{\em no} spreading.

\h On the analogy of conventional nonrelativistic quantum mechanics, one can
define from eq.(40) the mean time at which a photon passes through point
(or plane) $x$ as[48,105]:

\begin{equation} \langle t(x)\rangle = \int_{-\infty}^{\infty} t \,
J_{{\rm em},x}\drm t  = {{\int_{-\infty}^{\infty} t \, S_{x}(x,t)\drm t} \over
{\int_{-\infty}^{\infty} S_{x} (x,t)\drm t}}  \end{equation}  %%%eq.(41)

\noi (where, with the natural boundary conditions $\chi_{i}(0)= \chi_{i}(\infty)=0$,
we can use in the energy $E=\hbar c k$ representation the same time operator
already adopted for particles in nonrelativistic quantum mechanics; and hence
one can prove the equivalence of the calculations of $\langle t(x)\rangle$, \
$ \Drm \, t(x)$, \ etc., \  in both the time and energy representations).

\h In the case of fluxes which change their signs with time we can introduce
also for photons, following refs.[48,71], the quantities
$J_{{\rm em},x,{\pm}}  = J_{{\rm em},x} \; \Theta ({\pm} J_{{\rm em},x} )$
with the same physical meaning as for particles. Therefore, suitable
expressions for the mean values and variances of propagation,
tunnelling, transmission, penetration, and reflection durations can be
obtained in the same way as in the case of nonrelativistic
quantum mechanics for particles (just by replacing $J$ with $J_{\rm em}$).
In the particular case of quasi-monochromatic wavepackets, by using the
stationary-phase method (under the same boundary conditions considered in
Sect.2 for particles), we obtain for the photon {\em phase tunnelling time}
the expression

\begin{equation}   \tau^{\rm Ph}_{\rm tun,em} =\frac{2}{c \kappa_{\rm em}}
\end{equation} \ ,  %%%eq.(42)

\noi for $L \kappa_{\rm em} >>1$, quantity $L$ being the length of the
undersized waveguide (cf. Fig.3). \ Eq.(42) is to be compared with eq.(11a). \
From eq.(42) we can see that when $ L \kappa_{\rm em} > 2$, the effective
tunnelling velocity

\begin{equation}  \upsilon_{\rm tun,em}^{\rm Ph} = {{L} \over
{\tau^{\rm Ph}_{\rm tun,em}}} \end{equation}       %%%eq.(43)

\noi is Superluminal, i.e., larger than $c$. \ This result agrees with all the
known experimental results performed with microwaves (cf., e.g., refs.
[56,65,98]).\\

\section{Tunnelling times in frustrated total internal reflection
experiments}     %%%Sect.10

\h Some results of optical experiments with tunnelling photons were
described in ref.[60a], where it was considered the scheme here presented in
Fig.4a.  A light beam passes from a dielectric medium into an air slab with
width $a$. For incidence angles ${i}$ greater than the critical angle
$i_{\crm}$ of total internal reflection, most of the beam is reflected, and
a small part of it tunnels through the slab.  Here tunnelling occurs in the
$x$ direction, while the wavepacket goes on propagating in the $z$
direction. Its peak, which is emerging from the second interface, has
undergone a temporal shift, which is equal to the mean phase tunnelling
time $\langle \tau^{\rm Ph}_{\rm tun} $, and a spatial shift $D$ along $z$.
Since it is natural to assume that the propagation velocity $\upsilon_{z}$
along $y$ is uniform during tunnelling, then

\begin{equation}  D = \upsilon_{z} \langle \tau^{\rm Ph}_{\rm tun}
\rangle  \end{equation}  %%%eq.(44)

\noi so that the mean phase time can be simply obtained by measuring $D$.

\h Since tunnelling imposes also a change in the mean energy (or wavenumber)
of a wavepacket, and the plane wave components with smaller
incident angle are better transmitted than those with larger incidence angles,
then the emerging beam suffers an angular deviation $\delta i$, that can
be interpreted as a beam mean-direction rotation during tunnelling. And hence,
by taking into account formulas (12)-(12a) and Sect.4, we can conclude that
$\delta i$  and the quantity  $\langle \tau^{\rm La}_{z,{\rm tun}} \rangle =
(\Drm \, \tau_{\rm tun}^{Ph})^{1/2}$ are proportional to each other:

\begin{equation}  \delta i =\Omega \langle \tau^{\rm La}_{z,{\rm tun}}
\rangle  \end{equation}   %%%eq.(45)

\noi where $\Omega$ is the rotation frequency which was calculated in
ref.[60a]. So, the time $\langle \tau^{\rm La}_{z,{\rm tun}} \rangle =
(\Drm \, \tau_{\rm tun}^{Ph})^{1/2}$  too can be simply obtained by measuring
$\delta i$. \ Both these times characterize the intrinsic properties of the
tunnelling process, under the conditions imposed on the wavepackets (which
were described in Sect.2).

\h Let us remark that the conclusion of the authors of
ref.[60a] about the fact that the mean phase time $\langle
\tau^{\rm Ph}_{\rm tun} \rangle$ was inadequate as a definition of the
tunnelling time is not true, because they describe the wave function in
the air slab by the evanescent term $\exp{(-\kappa x)}$ only, instead of
considering the superposition $\alpha \exp{(-\kappa x)}+\beta
\exp{(\kappa x)}$ of {\em evanescent and anti-evanescent} waves. \ It is
important to recall that such a superposition of decreasing and increasing
waves ---normally used in the case of particle tunnelling--- is necessary
to obtain a resulting non-zero flux!.[48]

\h With such a correction, one can see that the very small values of
$\langle \tau^{\rm Ph}_{\rm tun} \rangle$ (about 40 fs) obtained in the
experiment[60a] for $a=20 \; \mu$ imply for the tunnelling photon a
{\em Superluminal peak velocity} of about $5 {\cdot} 10^{10}$ cm/s.

\h But in the double-prism arrangement, it was predicted by Newton, and
preliminarly confirmed 250 years later by F.Goos and H.H\"anchen, that
the reflected and transmitted beams are also spatially shifted with respect
to what expected from geometrical optics (cf. Fig.4b).  Recent rather
interesting esperiments have been performed by Haibel et al.[60c], who
discovered a strong dependence of the mentioned shift on the beam width
and especially on the incidence angle.

\section{A remark on reshaping} %%%Sect.11

\h The Superluminal phenomena, observed in the experiments with tunnelling
photons and evanescent electromagnetic waves[56-60], generated a lot of
discussions on relativistic causality[112-120].  This revived an interest
also in similar phenomena that had been previously observed in the case of
electromagnetic pulses propagating in dispersive media[88,121,122]. \
On the other side, it is well-known since long that the wavefront velocity
(well defined when the pulses have a step-function envelope or at least an
abruptly raising forward edge) cannot exceed the velocity of light $c$ in
vacuum[108,123]. \ Even more, the (Sommerfeld and Brillouin)
{\em precursors} ---that many people, even if not all, believe to be
necessarily generated together with any signal generation--- are known to
travel exactly at the speed $c$ in any media (for a recent approach to the
question, see ref.[124]). \ Such phenomena were confirmed by various
theoretical methods and in various processes, including
tunnelling[102,112-113,125]. \ Discussions are presently going on about the
question whether the signal velocity has to do with the previous speed $c$
of with the group velocity.[125,124]   Another point under discussion is
whether the shape of a realistic wavepacket must possess, or not, an abruptly
raising forward edge.[102,115-118].

\h A simple way of understanding the problem, in a ``causal" manner, might
consist in explaining the Superluminal phenomena during tunnelling as simply
due to a ``reshaping", with attenuation, of the pulse, as already
attempted (at the classical limit) in refs.[100-102]: namely, the later parts
of an incoming pulse are preferentially attenuated, in such a way that the
outcoming peak appears shifted towards earlier times even if it is nothing
but a portion of the incident pulse forward tail[57]. In particular,
the following scheme is compatible with the usual idea of causality: If the
overall pulse attenuation is very strong and, during tunnelling, the leading
edge of the pulse is less attenuated than the trailing edge, then the time
envelope of the outcoming (small) flux can stay totally beneath the initial
temporal envelope (i.e., the envelope of the initial pulse in the case of
free motion in vacuum).[116-120] \ And, if $A_{\rm T}$ depends on energy
much more weakly than the initial wavepacket weight-factor, then the spectral
expansion, {\em and hence the geometrical form}, of the transmitted
wavepacket will be practically the same as the spectral expansion, {\em and
the form}, of the entering wavepacket ({\em reshaping\/}). By contrast, if
the dependence of $A_{\rm T}$ on energy is not weak, the pulse form and width
can get strongly modified ({\em ``reconstruction"}).

\h The very definition of causality seems to be in need of some careful
revision[126]. Various, possible ({\em sufficient but not necessary)}
``causality conditions" have been actually proposed in the literature.
For our present purposes, let us mention that an acceptable, more general
causality condition (allowing the time envelope of the final flux,
$J_{\rm fin}$, to arrive at a point $x_f \geq a$ even earlier than that of
the initial pulse) might be for example the following one:

\begin{equation}  \int_{-\infty}^{t}  [J_{\rm in}(x_{f},\tau) -J_{{\rm fin},+} (x_{f},
\tau )] d\tau \geq 0 \; , \ \ \     -\infty < t <\infty \ ; x_f \geq a \ .
\end{equation}  %%%eq.(46)

\noi It simply requires that, during any (upper limited) time interval,
the {\em integral} final flux (along any direction) does not exceed the
{\em integral} ``initial" flux which {\em would} pass through the same
position $x_f$ in the case of
free motion; although one can find finite values of $t_{1}$ and $t_{2}$ \
($-\infty <t_{1} <t_{2} <\infty$) \ such that  \ $ \int_{t_{1}}^{t_{2}}
[J_{\rm in}(x_{f},\tau) - J_{{\rm fin},+}(x_{f},\tau)]d\tau < 0$.

\h But other conditions for causality can of course be proposed; namely:

$$ \frac{\int^{t_{0}}_{-\infty} t J_{{\rm fin},+}(x_{f},\tau)d\tau}
{\int^{t_{0}}_{-\infty} J_{{\rm fin},+}(x_{f},\tau)d\tau} -
\frac{\int^{t_{0}}_{-\infty} t J_{\rm in}(x_{f},\tau)d\tau}
{\int^{t_{0}}_{-\infty} J_{\rm in}(x_{f},\tau)d\tau} \geq 0 \ , \eqno(46a) $$

\noi where $t_{0}$ is the instant corresponding to the intersection
({\em after} the final-peak appearance) of the time envelopes of those
two fluxes. \ Relation (46a) simply means that there is a {\em delay}
in the (time averaged) appearance at a certain point $x_f > a$
of the forward part of the final wavepacket, with respect to the
(time averaged) appearance of the forward part of the initial wavepacket
in the case that it freely moved (in vacuum). \ Conditions (46) and (46a)
are rather general.

\h  It is curious that, without violating such causality conditions, a piece
of information, by means of a (low-frequency) modulation of a
(high-frequency) carrying wave, can be transmitted ---even if with a strong
attenuation--- with a {\em Superluminal} group velocity.\\
%\begin{mathletters}
%\label{generallebel}
%\begin{equation} \hat{t}=\left\{\begin{array}{ll} t & \mbox{in eigen($t-$)representation} \\ %\label{mlett:1}
% -i\hbar \frac{\partial}{\partial E} & \mbox{in energy($E-$)representation}  \end{array} \right. %\label{mlett:2}
%\end{equation}
%\end{mathletters}

\section{Tunnelling through successive barriers} %%%Sect.12

\h Let us finally study the phenomenon of one-dimensional
non-resonant tunnelling through two successive opaque potential
barriers[128], separated
by an intermediate free region $\Rcal$, by analyzing the relevant solutions
to the Schroedinger equation. \ We shall find that the total traversal time
does {\em not} depend not only on the opaque barrier widths (the so-called
``Hartman
effect"), but also on the $\Rcal$ width: so that the effective velocity in
the region $\Rcal$, between the two barriers, can be regarded as infinite. \
This agrees with the results known from the corresponding waveguide
experiments, which simulated the tunnelling experiment herein considered
due to the formal identity between the Schroedinger and the Helmholtz
equation.

\h Namely, in this Section we are going to show that ---when studying an
experimental setup with two rectangular opaque potential barriers (Fig.5)---
the (total) phase tunneling time through the two barriers does depend neither
on the barrier widths {\em nor on the distance between the barriers}.[128]

\h Let us consider the (quantum-mechanical) stationary solution for the
one-dimensional (1D) tunnelling of a non-relativistic particle, with mass $m$
and kinetic energy \ $E=\hbar^2k^2/2m=\frac{1}{2}mv^2$, \ through two equal
rectangular barriers with height $V_0$ ($V_0>E$) and width $a$, \
quantity $L-a \ge 0$ being the distance between them. \ The Schr\"{o}dinger
equation is

$$
-\frac{\hbar^2}{2m}\,\frac{\pa^2}{\pa x^2}\,\psi(x) \; + \; V(x)\,\psi(x)
\ug E\,\psi(x)\,,
\eqno{(47)}
$$

\noi where $V(x)$ is zero outside the barriers, while $V(x) = V_0$ inside the
potential barriers. In the various regions I~$(x\leq 0)$, II~$(0\leq x\leq a)$,
III~$(a\leq x\leq L)$, IV~$(L\leq x\leq L+a)$ and V $(x\geq L+a)$, the
stationary solutions to eq.(47) are the following

\

$\hfill{\displaystyle\left\{\begin{array}{l}
\psiuno \ug \erm^{+ikx} + \AR\,\erm^{-ikx}\\

\psidue \ug \al\,\erm^{-\chi x} + \be\,\erm^{+\chi x}\\

\psitre \ug \AT\,\left[\erm^{ikx} + \ARAR\,\erm^{-ikx}\right]\\

\psiquattro \ug \AT\,\left[\alal\,\erm^{-\chi(x-L)} +
\bebe\,\erm^{+\chi(x-L)}\right]\\

\psicinque \ug \AT\,\ATAT\,\erm^{ikx} \; ,
\end{array}\right.}
\hfill{\displaystyle\begin{array}{r}
(48{\rm a}) \\ (48{\rm b}) \\ (48{\rm c}) \\ (48{\rm d}) \\ (48{\rm e})\end{array}}$

\

\

\noi where $\chi\equiv\sqrt{2m(V_0-E)}/\hbar$, and quantities $\AR$, $\ARAR$,
$\AT$, $\ATAT$, $\al$, $\alal$, $\be$ and $\bebe$ are the reflection
amplitudes, the transmission amplitudes, and the coefficients of the
``evanescent" (decreasing) and ``anti-evanescent" (increasing) waves for
barriers 1 and 2, respectively. \ Such quantities can be easily obtained from
the matching (continuity) conditions:

\

$\hfill{\displaystyle\left\{\begin{array}{l}
\psiuno(0) \ug \psidue(0)\\
{\displaystyle\left.\frac{\pa\psiuno}{\pa x}\right|_{x=0}} \ug
{\displaystyle\left.\frac{\pa\psidue}{\pa x}\right|_{x=0}}
\end{array}\right.}
\hfill{\displaystyle\begin{array}{r}
(49{\rm a}) \\ (49{\rm b})\end{array}}$

\

\

$\hfill{\displaystyle\left\{\begin{array}{l}
\psidue(a) \ug \psitre(a)\\
{\displaystyle\left.\frac{\pa\psidue}{\pa x}\right|_{x=a}} \ug
{\displaystyle\left.\frac{\pa\psitre}{\pa x}\right|_{x=a}}
\end{array}\right.}
\hfill{\displaystyle\begin{array}{r}
(50{\rm a}) \\ (50{\rm b})\end{array}}$

\

\

$\hfill{\displaystyle\left\{\begin{array}{l}
\psitre(L) \ug \psiquattro(L)\\
{\displaystyle\left.\frac{\pa\psitre}{\pa x}\right|_{x=L}} \ug
{\displaystyle\left.\frac{\pa\psiquattro}{\pa x}\right|_{x=L}}
\end{array}\right.}
\hfill{\displaystyle\begin{array}{r}
(51{\rm a}) \\ (51{\rm b})\end{array}}$

\

\

$\hfill{\displaystyle\left\{\begin{array}{l}
\psiquattro(L+a) \ug \psicinque(L+a)\\
{\displaystyle\left.\frac{\pa\psiquattro}{\pa x}\right|_{x=L+a}} \ug
{\displaystyle\left.\frac{\pa\psicinque}{\pa x}\right|_{x=L+a}}
\end{array}\right.}
\hfill{\displaystyle\begin{array}{r}
(52{\rm a}) \\ (52{\rm b})\end{array}}$

\

\

\h Equations (49-52) are eight equations for our eight unknowns ($\AR$, $\ARAR$,
$\AT$, $\ATAT$, $\al$, $\alal$, $\be$ and $\bebe$). \ First, let us obtain
the four unknowns $\ARAR$, $\ATAT$, $\alal$, $\bebe$ from eqs.(51) and (52)
in the case of {\em opaque} barriers, i.e., when $\chi a\rightarrow\infty$:

\

$\hfill{\displaystyle\left\{\begin{array}{lcr}
\alal \longrightarrow \erm^{ikL}\,\displaystyle{\frac{2ik}{ik-\chi}}& \
\hskip 4cm & (53{\rm a})\\

\bebe \longrightarrow \erm^{ikL-2\chi a}\,\displaystyle{\frac{-2ik(ik+\chi)}
{(ik-\chi)^2}}& \ & (53{\rm b})\\

\ARAR \longrightarrow \erm^{2ikL}\,\displaystyle{\frac{ik+\chi}{ik-\chi}}& \
& (53{\rm c})\\

\ATAT \longrightarrow \erm^{-\chi a}\erm^{-ika}\,\displaystyle
{\frac{-4ik\chi}{(ik-\chi)^2}} & \ & (53{\rm d})
\end{array}\right.}$

\

\h Then, we may obtain the other four unknowns $\AR$, $\AT$, $\al$, $\be$ from
eqs.(49) and (50), again in the case $\chi a\rightarrow\infty$; \ one gets
for instance that:
$$
\AT \ug - \erm^{-2\chi a} \; \displaystyle{{4 i \chi k} \over {(\chi - i k)^2}}
 \ A
\eqno{(54)}
$$

\noi where
$$
A \equiv \displaystyle{{2 \chi k} \over {2 \chi k \; \cos k(L-a) \, + \,
(\chi^2 - k^2) \; \sin k(L-a)}}
$$

\

\noi results to be {\em real}; and where, it must be stressed,

\

$$
\delta\equiv\arg\left(\frac{ik+\chi}{ik-\chi}\right)
$$

\

\noi is a quantity that {\em does not depend} on $a$ and on $L$. \ This is
enough for concluding that the phase tunnelling time (see, for instance,
refs.[42,48,66,67,70,128])

$$
{\tau^{\rm ph}}_{\rm tun} \equiv
\hbar\,\frac{\pa\arg\left[\AT\ATAT\erm^{ik(L+a)}\right]}{\pa E} \; =  \;
\hbar \, {{\pa} \over {\pa E}} \, \arg{\left[{{-4ik\chi} \over {(ik-\chi)^2}}
\right]} \ ,
\eqno{(55)}
$$

\noi while depending on the energy of the tunnelling particle, {\em does not
depend} on $L+a$ (it being actually independent both of $a$ and of $L$).

\h This result does {\em not only} confirm the so-called ``Hartman
effect"[48,66,67,70,128] for the two opaque barriers ---i.e., the
independence of the tunnelling
time from the opaque barrier widths---, but it does {\em also} extend 
such an effect by implying the total tunnelling time to be independent even 
of $L$ (see Fig.5): something that may be regarded as a further evidence of
the fact that quantum systems seem to behave as
non-local.[128-131,88,87,71,48] \  It is important
to stress, however, that the previous result holds only for non-resonant (nr)
tunnelling: \ i.e., for energies far from the resonances that arise in region
III due to interference between forward and backward travelling waves (a
phenomemon quite analogous to the Fabry-P\'erot one in the case of classical
waves). Otherwise it is known that the general expression for (any) time
delay $\tau$ near a resonance at $E_\rrm$ with half-width $\Ga$ would be \
$\tau \: = \: \hbar \Ga [(E - E_\rrm)^2 + \Ga^2]^{-1} \: + \: \tau_{\rm nr}$. 

\h The tunnelling-time independence from the width ($a$) of each one of the two
opaque barriers is itself a generalization of the Hartman effect, and can be a
priori understood ---following refs.[57,62] (see also refs.[64,55])--- on the
basis of the reshaping phenomenon which takes place inside a barrier.

\h With regard to the even more interesting tunnelling-time independence from the
distance $L-a$ between the two barriers, it can be understood on the basis of
the interference {\em between the waves} outcoming from the first barrier
(region II) and traveling in region III {\em and the waves} reflected from
the second barrier (region IV) back into the same region III. \ Such an
interference has been shown[48,70,131] to cause an ``advance" (i.e.,
an effective
acceleration) on the incoming waves; a phenomenon similar to the analogous
{\em advance} expected even in region I.  Namely, going on to the
wavepacket language, we noticed in ref.[48,70,131] that the arriving wavepacket
does interfere with the reflected waves that start to be generated as soon
as the packet forward tail reaches the (first) barrier edge: in such a way
that (already before the barrier) the backward tail of the initial wavepacket
decreases ---for destructive interference with those reflected waves--- at a
larger degree than the forward one. This simulates an increase of the average
speed of the entering packet: hence, the effective (average) flight-time
of the approaching packet from the source to the barrier does decrease.

\h So, the phenomena of reshaping and advance (inside the barriers and
to the left of the barriers) can qualitatively explain why the
tunnelling-time is independent of the barrier widths and of the distance
between the two barriers. \ It remains impressive, nevertheless, that
in region III ---where no potential barrier is present, the current  is
non-zero and the wavefunction is oscillatory,--- the effective speed
(or group velocity) is practically {\em infinite}. Loosely speaking, one 
might say that the considerd two-barriers setup is an ``(intermediate) 
space destroyer". \ After some straightforward but rather bulky calculations, 
one can moreover see that the same effects (i.e., the independence from the 
barrier widths and from the distances between the barriers) are still valid 
for any number of barriers, with different widths and different distances 
between them.

\h Finally, let us mention that the known similarity between photon and
(nonrelativistic) particle tunneling[48,57,61,62,70,132; see also 55,64,130]
implies our previous results to
hold also for photon tunnelling through successive ``barriers": For example,
for photons in presence of two successive band gap filters: like two suitable
gratings or two photonic crystals.  Experiments should be easily realizable;
while indirect experimental evidence seems to come from papers as [129,121].

\h At the classical
limit, the (stationary) Helmholtz equation for an electromagnetic
wavepacket in a waveguide is known to be mathematically identical to the
(stationary) Schroedinger equation for a potential barrier;\footnote{These
equations are however different (due to the different order
of the time derivative) in the time-dependent case. Nevertheless, it can be
shown that they still have in common classes of analogous solutions,
differing only in their spreading properties[48,70,61, and 131].} \ so that,
for instance, the tunnelling of a particle through and under a barrier can be
simulated[48,70,58-62,86,132-134] by the traveling of evanescent waves along an undersized
waveguide. Therefore, the results of this paper are to be valid also for
electromagnetic wave propagation along waveguides with a succession of
undersized segments (the ``barriers") and of normal-sized segments. This
agrees with calculations performed, within the classical realm, directly
from Maxwell equations[130,86,134,135], and has already been confirmed
by a series of ``tunnelling" experiments with microwaves: see
refs.[58-60,133] and particularly [116,136].\\

\section{Conclusions and prospects}      %%%Sect.13

\h  {\bf I}. A basic physical formalism for determining the collision
and tunnelling times for nonrelativistic particles and for photons seems
to be now available:

(1) We have found selfconsistent definitions for the mean times and
durations of various collision processes (including tunnelling), together
with the variances of their distributions. This was achieved by utilizing
the properties of {\em time}, regarded as a quantum observable (in quantum
mechanics and in quantum electrodynamics).

(2) Such definitions seem to work rather well, at least for large
(asymptotic) distances between initial wavepackets interaction
region, and for finite distances between interaction region and final
wavepackets. In these cases the phase-time, the clock and the instanton
approaches yield results which happen to coincide either with the mean
duration or with the standard deviation [square root of the
duration-distribution variance] forwarded by our own formalism. And the
(asymptotic) mean dwell time results to be the average weighted sum of the
tunnelling and reflection durations: cf. eq.(16).

\h Notice that formulae (4), (6), (8) can be rewritten in a unified way
(in terms of the mean square time durations) as follows:

$$ \langle[ \tau_{\Nrm} (x_{i}, x_{f})]^{2} \rangle =
[ \langle \tau_{\Nrm} (x_{i}, x_{f}) \rangle ]^{2}+  \Drm \tau_{\Nrm} (x_{i}, x_{f})
\eqno(56) $$

\noi with $\Drm \tau_{\Nrm} (x_{i}, x_{f}) = \Drm \, t_{\srm} (x_{f}) +
\Drm \, t_{+} (x_{i})$, where $N$ may mean ${\rm T}$ or ${\rm tun}$
or ${\rm pen}$ or ${\rm R}$, etc., and $\srm = +,-$: more precisely,
$\srm = -$ in the case of reflection and $\srm = +$ in the remaining cases. \
Relations (56) can be further on rewritten in the following equivalent forms:

$$ \langle[ \tau_{\Nrm} (x_{i}, x_{f})]^{2} \rangle = [ \langle
t_{\srm}(x_{f}) -
t_{+} (x_{i}) \rangle ]^{2}+ \Drm t_{\srm}(x_{i}) \ . \eqno(56a) $$

\noi We can now see that the square phase duration $[ \langle
\tau_{\rm T}^{\rm Ph} \rangle ]^{2}+\Drm \, \tau_{\rm T}^{\rm Ph}$,
and the square hybrid time $[ \tau_{y,{\rm tun}}^{\rm La})^{2} +
(\tau_{z,{\rm tun}}^{\rm La})^{2}]^{2}$ introduced by B\"{u}ttiker[79],
as well as the square magnitude of the complex tunnelling time in the
Feynman path-integration approach, are all {\em examples of mean square
durations}.  Notice that the Feynman approach (in the case of its instanton
version) and the B\"{u}ttiker hybrid time (in the case of an infinite
extension of the magnetic field) coincide with the mean square phase
duration.\hfill\break

\h  By the way, our present formalism has been already applied and tested in
the time analysis of nuclear and atomic collisions for which the boundary
conditions are experimentally and theoretically assigned in the region,
asymptotically distant from the interaction region, where the incident
(before collision) and final (after collision) fluxes are well separated in
time, without any superposition and interference.  And it has been supported
by results (see, in particular, refs.[29,30] and references therein)
such as:

(i) the validity of a correspondence principle between the time-energy
QM commutation relation and the CM Poisson brackets;

(ii) the validity of an Ehrenfest principle for the average time durations;

(iii) the coincidence of the quasi-classical limit of our own QM definitions
for time durations (when such a limit exists; i.e. for above-barrier
energies) with analogous well-known expressions of classical mechanics;

(iv) the direct and indirect experimental data on nuclear-reaction durations,
in the range $10^{-21} \div 10^{-15}$ s, and the compound-nucleus level
densities extracted from those data.

\h  Let us mention that for a complete extraction of the time-durations from
{\em indirect} measurements of nuclear-reaction durations it is necessary to
have at disposal correct definitions not only of the mean durations but also
of the duration variances[30], as provided by our formalism.

\h At last, let us recall that such a formalism provided also useful tools
for resolving some long-standing problems related to the time-energy
uncertainty relation[29,30].
%; and that some recent theoretical work by
%Abolhasani and Golshani[71], which regarded our OR approach as giving
%the most natural definition for a transmission time
%within the standard interpretation of quantum mechanics, conluded that
%the best times that could be obtained in Bohmian mechanics are {\em the
%same} as OR's.\\

\h  {\bf II.}  In order to apply the present formalism to the cases when one
considers not only asymptotic distances, but also the region inside and near
the interaction volume, we had to revise the notion of weighted average
(or integration measure) in the time representation, by adopting the two
weights $J_{\pm}(x,t) \drm t$ when evaluating instant and duration mean
values, variances, etc., for a moving particle, and the third weight
$dP(x,t)$ or $P(x_{1},x_{2};t) \drm t$ when calculating mean durations for
a ``dwelling" particle.  And in terms of these three weights we can express
all the different approaches proposed within conventional quantum mechanics,
including the mean dwell time, the Larmor-clock times, and the times
given by the various versions of the Feynman path-integration approach:
Namely, we can put them all into a single {\em non-contradictory} scheme
on the basis of our formalism, even for a particle inside the barrier.

\h The same three weights can be used also in the analogous
quantum-mechanical formalism for the {\em space} analysis of collision and
propagation processes (see also [18]).\\

\h  {\bf III.}  Our flux separation into $J_+$ and $J_-$ is not the only
procedure to be possible within {\em conventional quantum mechanis (and
quantum electrodynamics)}, although it is the only {\em non-coherent} flux
separation known to us avoiding the introducing of any new postulates.
In fact, one can also adopt the ``{\em coherent} wavepacket separation"
into positive and negative momenta, which has a clear meaning outside the
barrier, but is obtainable only via a mathematical tool like the momentum
Fourier expansion inside the barrier. Such a separation can be transformed
into an ``incoherent flux separation" by exploiting the postulate of the
measurement quantum theory about the possibility of describing the
measurement conditions in terms of the corresponding projectors: that is
to say, of the projectors $\Lambda_{\exp,{\pm}}$ onto positive-momentum
and negative-momentum states, respectively [cf. eq.(13a), Sect.2]. \ There
are also flux separation
schemes within nonstandard versions of quantum theory (cf., e.g., Sect.5). \
However, whatever separation scheme we choose, we have to stick to at least
two necessary conditions:

(A) each normalized flux component must possess a probabilistic meaning, and

(B) the standard flux expressions, well-known in quantum collision theory,
must be recovered in the asymptotically remote spatial regions.\hfill\break

\h In brief, with regard to the region inside and near a barrier, at least
four kinds of separation procedures for the wavepacket fluxes do exist, which
satisfy the previous conditions:

(i) The OR separation  $J = J_{+} + J_{-}$, with
$J_{\pm} = J \; \Theta ({\pm} J )$, which was obtained from the 
conventional continuity equation for probability (i.e., from the
time-dependent Schroedinger equation) without any new physical postulates
or any new mathematical approximations[71]. The asymptotic behaviour, e.g.,
of the obtained expressions was tested by comparison with other approaches
and with the experimental results[48]; see also point (v) below.

(ii) The separation proposed here, i.e., $J=J_{\exp,+} + J_{\exp,-}$
(quantities $J_{\exp,{\pm}}$ being the fluxes which correspond to
$\Lambda_{\exp,{\pm}} (x,t)$, respectively), is also a consequence of the
conventional probability continuity equation, provided that it is accepted
the wave-function reduction postulate of ordinary quantum measurement theory. \
It corresponds to the adoption of ``semi-permeable" detectors, which are open
for particles arriving from one direction only. \ The asymptotic behaviour
of the expressions, obtained on the basis of this separation, coincides with
that yielded by (i).

(iii) Relation (20) was obtained in the Muga-Brouard-Sala approach, within
the physically clear ``incoherent flux separation" of positive and negative
momenta, but with the additional introduction of the Wigner-path
distributions.

(iv) Relation (26) was obtained in the Leavens' approach, on the basis of
an incoherent flux separation of the trajectories of particles to be
transmitted from particles to be reflected, via the introduction of
the nonstandard Bohm interpretation of quantum mechanics.

\h    The flux separation schemes (i), (iii) and (iv) yield asymmetric
expressions for the mean dwell time near a barrier [equations (15), (21) and
(22)-(25), respectively], apparently due to the right-left asymmetry of the
boundary conditions: we have incident and reflected wavepackets on the
left, and only a transmitted wavepacket on the right. \ The separation
procedure (ii) yields the symmetric expression (16) for the mean dwell time
even near a barrier.\\

\h  {\bf IV.} In Sect.7 we have shown that (in the absence of absorption and
dissipation) the Hartman effect is valid for {\em all} the mean tunnelling
times, while it does not hold only for the quantities that at a closer
analysis did not result to be tunnelling times, but rather tunnelling-time
{\em standard deviations}.

\h Let us recall at this point that only the sum of increasing (evanescent)
and decreasing (anti-evanescent) waves corresponds to a non-zero
{\em stationary} flux. Considering such a sum is standard in quantum
mechanics, but not when studying evanescent waves (the analogue of tunnelling
photons) in classical physics. On the contrary, that sum should of course be
taken into account also in the latter case, obtaining non-zero
(stationary) fluxes.

\h In any case, it is interesting to notice that in the {\em non-stationary}
case, even evanescent waves alone, or anti-evanescent waves alone, correspond
separately to {\em non-zero} fluxes.  Even more, from the general expression
of a non-stationary wave packets inside a barrier, one can directly see that,
e.g., {\em evanescent} waves (considered alone) seem to fill up
instantaneously the entire barrier as a whole!; this being a further evidence
of the {\em  non-local} phenomena which take place during {\em sub-barrier}
tunnelling. \ Even stronger examples of non-locality have been met by us
in Sect.12 above: cf. eq.(55). \ Some numerical evaluations[86,98,126],
based on Maxwell equations only, showed that analogous phenomena occur for
classical evanescent waves in under-sized waveguides (``barriers"), as
confirmed by experience.  /  Let us recall, at last, that even Superluminal
localized (non-dispersive, wavelet-type) pulses which are solutions to the
Maxwell equations
have been constructed[130], which are {\em not} evanescent but on the
contrary propagate without distortion along {\em normal} waveguides.\\

\h  {\bf V.} In connection with Sects.2, 6 and 11, let us recall that the
requirement that the values of the collision, propagation, tunnelling
duration be positive is a {\em sufficient} but not necessary causality
condition.  Therefore we have not got a unique general formulation
of the causality principle which is necessary for all possible cases.
In Sect.2 and 11 some new formulations of the causality condition heve
been by us just proposed.\\

\h  {\bf VI.} The phenomena of reshaping, which were dealt with in Sect.9,
as well as the ``advance" which takes place before the barrier entrance
(discussed in Sect.6) are closely connected with the (coherent) superposition
of incoming and reflected waves.  \ Moreover, the study of reshaping (or
reconstruction) and of the advance phenomenon can be of help, by
themselves, in understanding the problems connected with
Superluminal phenomena and the definition of signal velocity[115-120].\\

\h  {\bf VII.} In the case of tunnelling through two successive opaque
barriers (cf. Fig.5), we strongly generalized the Hartman effect, by showing
in Sect.12 that {\em far from resonances} the (total) phase tunneling time
through the two opaque barriers ---while depending on the energy--- is {\em
independent} not only of the barrier widths, but even  {\em of the distance
between the barriers}: \ So that the effective velocity in the free
region, between the two barriers, can be regarded as infinite.\\

\h  {\bf VIII.} We mentioned in Sect.8 that the two-phase  description of
tunnelling can be convenient for media without absorption and
dissipation, and also for Josephson junctions.\\

%%%mancano 127, 137, 138, 139, 140

\h   {\bf IX.}
%%%Another approach was based on a (slightly
%%%modified) instanton method[139], or on a combination of the instanton and of
%%%the deformed space metric inside the barrier[140].\\
The OR formalism, as presented in this paper, permits in principle to study
the time evolution of collisions in the Schroedinger and Feynman
representations (which lead, by the way, to the same results). An interesting
attempt was undertaken in ref.[141] to a selfconsistent description of a
particle motion, by utilizing the Feynman representation and comparing their
method with the OR formalism (in its earlier version, presented in
ref.[48]), even if skipping the separation $J=J_{+} + J_{-}$ .

\h There is one more possible representation, equivalent to Schroedinger's
and Feynman's, for investigating the collision and tunnelling evolution. Let
us recall that in quantum theory to the {\em energy} $E$ there correspond
the two operators $i\hbar \partial /\partial t$ and the hamiltonian operator.
Their duality is well represented by the Schroedinger equation
$H \Psi = i \hbar \partial \Psi /\partial t$. \ A similar duality does
exist in quantum mechanics for {\em time\/}: besides the general form
$-i \hbar \partial /\partial E$, which is valid for any physical systems
(in the continuum energy spectrum case), it is possible to express the time
operator $\hat{T}$ (which is hermitian, and also maximal hermitian[27,18,19],
even if not self-adjioint) in terms of the coordinate and momentum
operators[25,30,142,143], by utilizing the commutation relation
$[\hat{T},\hat{H}] = i \hbar $. \ So that one can study the collision and
tunnelling evolutions via the operator $\hat{T}$ by the analogous equation
$\hat{T} \Psi = t \Psi$, particularly for studying he influence of the
barrier shape on the tunnelling time[30].\\

\h   {\bf X.} In Sect.9 the analogy between tunnelling processes of
photons (in first quantization) and of non-relativistic particles has been
discussed and clarified, and it was moreover shown that the properties
of time as an observable can be extended from quantum mechanics to
one-dimensional quantum electrodynamics. On the basis of this analogy,
in Sects.9 and 10 a selfconsistent interpretation of the photon
tunnelling experiments, described in refs.[56,60], was presented.\\

\h    {\bf XI.} At last, let us mention that for discrete energy spectra,
the time analysis of the
processes (and, particularly, in the case of wavepackets composed of
states bound by two well potentials, with a barrier between the wells) is
rather diffeent from the time analysis of processes in the continuous energy
spectra. For the former, one may use the formalism[30,31] for the time
operator in correspondence with a discrete energy spectrum: and the durations
of the transitions from one well to the other happen to be given by the
Poincar\'{e} period $2 \pi \hbar /d_{\minrm}$, where $d_{\minrm}$ is the
highest common factor of the level distances, which is determined by the
{\em minimal level splitting caused by the barrier} and hence depends
on the barrier traversal probability at the relevant energies[144].

\h  One can expect that the time analysis of more complicated processes,
in the quasi-discrete (resonance) energy regions, with two (or more)
well-potentials, such as the {\em photon ot phonon-induced tunnellings from
one well to the other}, could be performed by a suitable combination
and generalization of the methods elaborated for continuous and discrete
energy spectra.\\

\

{\bf Acknowledgements}\\

For discussions, the authors are grateful to J.Brown, G.Salesi, G.Privitera,
A.S.Holevo, J.Le\'on, B.Mielnik, E.Kapu\'scik, A,Shaarawi, A.Ranfagni,
R.Chiao, F.Fontana, B.N.Zakhariev, V.L.Lyuboshitz, A.Agresti, D.Mugnai,
A.K.Zaichenko, R.Landauer, M.B\"uttiker, G.Kuritzi, E.H.Hauge, S.A.Omelchenko,
S.P.Maidanyuk, D.Ahluwalia, A.Fass\`{o}, F.Fontana, R.Bonifacio, G.Degli
Antoni, H.E.Hern\'andez-Figueroa, J.Swart, A.Steinberg, J.Muga, F.Bassani,
R.A.Ricci, R.H.Farias, J.Vaz, Wagner, E.Conforti, A.Salanti, R.Riva,
R.Colombi, M.Zamboni-Rached, G.Nimtz, J.-y.Lu, A.Bertin.

\

\

\end{document}